\documentclass[12pt]{article}
\usepackage{hyperref}
\usepackage{amsmath}
 \usepackage{bbm}
\usepackage{cite}
\newsavebox{\myhbar}
\savebox{\myhbar}{$\hbar$}

\usepackage{graphicx}
\usepackage{enumerate}
\usepackage{mdframed}
\usepackage{braket}
\usepackage{url}
\usepackage{wasysym}

\usepackage[hmargin=2.5cm,vmargin=3cm]{geometry}
\usepackage{subfig}
\usepackage{graphicx}
\usepackage{float}
\renewcommand*{\hbar}{\mathalpha{\usebox{\myhbar}}}

\newcommand{\bey}{\begin{eqnarray}}
\newcommand{\eey}{\end{eqnarray}}

\begin{document}

 \title{Thermodynamics and phase transitions of black holes in contact with a gravitating heat bath}
  \author { Demetrios Kotopoulis\footnote{d.kotopoulis@upatras.gr} \; and \;
 Charis Anastopoulos\footnote{anastop@upatras.gr}\\
 Department of Physics, University of Patras, 26500 Greece}
\maketitle

\maketitle

\begin{abstract}
We study the thermodynamics of a shell of self-gravitating radiation, bounded by two spherical surfaces. This system provides a consistent model for a gravitating thermal reservoir for different solutions to vacuum Einstein equations in the shell's interior. The latter include black holes and flat space, hence, this model allows for the study of black hole phase transitions.
 Following the analysis of  {\em Class. Quant. Grav. 29, 025004 (2012)}, we show that the inclusion  of appropriate entropy terms to the spacetime boundaries (including the Bekenstein-Hawking entropy for black hole horizons) leads to a consistent thermodynamic description. The system
 is characterized by four phases, two black hole phases distinguished by the size of the horizon, a flat space phase and one phase that describes naked singularities. We undertake a detailed analysis of  black-hole phase transitions, the non-concave entropy function, the properties of temperature at infinity, and system's heat capacity.
\end{abstract}

\pagebreak

\section{Introduction}
Ever since Bekenstein's proposal of black hole entropy \cite{Bek1}  and Hawking's derivation of black hole radiation \cite{Hawk} , black holes are understood as thermodynamic objects. According to the generalized second law of thermodynamics (GSL)\cite{Bek2}, black hole entropy adds up with the entropy of matter, and their sum is a non-decreasing function of time.

The GSL implies that there exists a larger thermodynamic space that contains black holes and self-gravitating systems,  i.e., systems of ordinary matter in which the gravitational self-interaction contributes significantly to their thermodynamic properties. Therefore, we expect the presence of
phase transitions between black holes and  self-gravitating systems.

In this work, we undertake the analysis of such phase transitions in a simple system, that consists of a shell of self-gravitating radiation. The geometry inside the shell  is  either of Minkowski space, or of a black hole or of a singular solution. The different interior geometries correspond to different thermodynamic phases for the system, while the shell acts as a thermal bath for these phases. We analyze the thermodynamic properties of the system and the characteristics of the associated phase transitions.


\subsection{Motivation}

There is substantial work on black-hole to black-hole phase transitions, first on Kerr-Newman black holes \cite{Davies, Hut, Rupeiner}, and later on asymptotically anti-de Sitter (AdS) black holes \cite{CCK00, KM12, AKMS14, KMT17}. Phase transitions in self-gravitating systems have also been extensively studied, see, for example, \cite{Cha02, Cha05, TVBR, CVB}. However, there is much less work on phase transitions between black holes and self-gravitating systems. The most well known case is the Hawking-Page phase transition between black holes and radiation \cite{HaPa}, albeit in asymptotically AdS spacetimes. In asymptotically flat spacetimes,  phase transitions have been studied by comparing the entropy of a Schwarzschild black hole in a box with the entropy of (non-gravitating) radiation in the box \cite{Davies, Hut, Hawk76, Page2, York}.  Backreaction from the Hawking radiation can also be included in the thermodynamical description \cite{AnSav16}.

 Certainly,  understanding phase transitions between black holes and self-gravitating systems is by itself important. Furthermore, progress in this direction could provide significant insights to quantum gravity theories \cite{Hu10}. For example, consider
a self-gravitating system that involves conserved quantities like particle numbers. These quantities disappear in the black hole case, by virtue of the no-hair theorem. A thermodynamic study of   the formation of  a black hole phase
can provide novel information about the relation between black hole hair and quantum effects, the latter being incorporated in black hole entropy. Thermodynamic processes that cross from the black hole phase to ordinary matter may be relevant to discussions of black hole information loss. In the longer term, an analysis of critical exponents, or of non-equilibrium properties in such phase transitions could  constrain  candidate quantum theories of gravity.

A different context for this work is the long-standing effort to understand the physics of non-extensive thermodynamic systems in equilibrium. Non-extensivity arises whenever the range of interactions of the system is larger than the size of the system: this is possible, either with short-range forces in small systems \cite{Gross} or  with long-range forces  \cite{DRAW,CDR09}, including gravity \cite{Pad90, Katz03}. Non-extensive systems exhibit  novel thermodynamic properties. They are  spatially inhomogeneous even in equilibrium, their micro-canonical and  canonical ensembles are inequivalent \cite{Thirring}, their entropy function is not be  concave (hence, heat capacities may be negative).
The model examined here is novel and informative for non-extensive thermodynamics. It involves full General Relativity (rather than Newtonian approximations), it brings together black holes and self-gravitating system. Crucially, it can be treated semi-analytically, thereby,   providing a workable testing ground for understanding the structure of gravitational thermodynamics.

\subsection{Background}

The natural place to start an analysis of black hole phase transitions is  the simplest
 self-gravitating system in  General Relativity, namely, static, spherically symmetric solutions to Einstein's equations. These are described by the
  Tolman-Oppenheimer-Volkoff (TOV) equation. One might expect that for a given type of matter, there may exist a region of the parameter space that corresponds to black hole horizons coexisting with matter. This turns out to not be the case: if matter satisfies the dominant energy conditions, then no horizons are encountered when integrating the TOV equation from the boundary inwards \cite{ST, AnSav21}. In fact, if the equation of state is thermodynamically consistent, there are only two types of solutions, regular ones and singular ones \cite{AnSav21}. The former are everywhere locally Minkowskian, and they describe ordinary compact stars.
  The latter contain a strongly repulsive naked singularity at the center, where the geometry is locally isomorphic to negative-mass
Schwarzschild spacetime.

The situation is different when we consider a {\em shell} rather than a {\em ball} of matter. Suppose we place an interior boundary at $r = r_0$ so that the geometry for $r < r_0$ is a solution to the vacuum Einstein equations. Then, the associated thermodynamic state space contains regular solutions, singular solutions and also black hole solutions. Hence, we can undertake a thermodynamic analysis that includes black hole phase transitions.

In this paper, we study a shell of radiation, bounded between two reflecting and non-thermally conducting surfaces at $r = r_0$ and at $r = R$. The simplicity of the equation of state, implies that some properties of the solutions can be evaluated semi-analytically.
For past studies of self-gravitating radiation, see, Refs. \cite{PWZ, ZuPa84, PL, AnSav12, Kim17, Kim17b}.

The radiation shell considered here defines a concrete model of a  thermal reservoir in contact with  a black hole. This model is an improvement of existing ones---see, for example,  \cite{York}---which employ the rather extreme idealization of switching off the gravitational interaction of the reservoir. Such an approximation is justified in extensive systems: when a system is brought into contact with a reservoir, it may acquire the reservoir's temperature or pressure, but its constitutive equations remain unchanged. This is due to the fact that short range forces affect only a small region around the interface of the system with the reservoir. In contrast, when long-range forces are involved, the reservoir acts directly on the whole of the system, and it can affect its constitutive equation.

Our analysis employs the formal structure of equilibrium thermodynamics, as described by Callen \cite{Callen}. We mentioned earlier that there exists as yet no set of thermodynamic axioms applicable to non-extensive systems. In principle, some axiomatic approaches to thermodynamics  \cite{Carath, Giles, LiYv1} can be generalized for non-extensive systems; for work in this direction, see Ref.  \cite{LiYv2}.
Still, a definitive axiomatic formulation of thermodynamics for gravitating systems does not yet exist. We note the work of Martinez in adapting
Callen's axioms   to gravitational systems in Ref. \cite{Martinez}, and Ref. \cite{AnSav14} for formulating the three laws of thermodynamics   in this context.

The key point in Callen's formulation of thermodynamics is that an isolated thermodynamic system is described in terms of a set of macroscopic constraints that determine its thermodynamic state space $Q$. The values of any variable that is not fixed by the constraints correspond to global maxima of the entropy function, subject to the constraints. This statement is the Maximum Entropy Principle (MEP). Eventually, all thermodynamic information is contained in the {\em fundamental thermodynamic function}, i.e., the entropy function $S: Q \rightarrow R^+$. In the present system, the thermodynamic state space $Q$ is three dimensional, it is determined by the shell radii $r_0$ and $R$ and by the Arnowitt-Deser-Misner (ADM) mass $M$.

\subsection{Results}

Our results are the following.

\medskip

\noindent $1.$ The space $Z$ of solutions to the TOV equation is larger than the thermodynamic state space $Q$. This means that the equilibrium values of the additional degrees of freedom must determined the MEP. We show that if the entropy function involves only contributions from radiation, the entropy function is unbounded and the MEP cannot be implemented. The system would then have no consistent thermodynamic description. This is an analogue of the {\em gravothermal catastrophe} \cite{gravothermal, gravothermal2}  that appears in non-relativistic self-gravitating systems.

\smallskip

\noindent $2.$ To restore thermodynamic consistency, we must add entropy terms associated to the internal boundaries of the system, i.e., the horizons and the singularities that appear in the interior region. For horizons, we use the Bekenstein-Hawking entropy. For the repulsive singularity, the associated entropy is obtained  by Wald's Noether charge for spacetime boundaries \cite{Wald93}, modulo a multiplicative constant. The latter is determined {\em uniquely} by the requirement of thermodynamic consistency, i.e., that the MEP can be implemented. The resulting expression for the boundary entropy is consistent with a previous result of Ref. \cite{AnSav12} for a ball of self-gravitating radiation.

\smallskip

\noindent $3.$ After the implementation of the MEP, we construct the entropy function on the thermodynamic state space $Q$. $Q$ splits into four regions, corresponding to four distinct phases. Phase F corresponds to a shell in locally Minkowski spacetime, phase $B_I$ describes a large black hole solution with little radiation in the shell, phase $B_{II}$ describes a small black hole coexisting with radiation, and phase S corresponds to singular solutions.

\smallskip

\noindent $4.$ We find that the phase transitions between the F phase and the $B_I$ and  $B_{II}$ phases are first-order. The latent heat in all transitions from the F phase to the black hole phases is negative, i.e., a heat must be removed from a self-gravitating system in order to form a black hole. We explain that this occurs because black holes constitute  a higher-energy phase but {\em not} a higher-temperature phase.
 The transition between the $B_I$ and the $B_{II}$ phases is continuous, and so is the transition between the F and S phases. There is also one {\em triple point} for the F, $B_I$ and $B_{II}$ phases.

\smallskip

\noindent $5.$ There is no coexistence curve between the black hole phases and the $S$ phase. This could change by a modification of the Bekenstein-Hawking expression for black hole entropy at very small masses.

\smallskip

\noindent $6.$ We analyse the behavior of other thermodynamic observables, focusing on the temperature at infinity,  on the heat capacity. The heat capacity of the system can become negative, however, in the context of non-extensive thermodynamics, this is not necessarily a sign of instability \cite{Velasquez}.

\medskip

The plan of this paper is the following. In Sec. 2. we analyse the geometry of the shell of  gravitating radiation, and all types of solutions. In Sec. 3, we show that the inclusion of entropy contributions from boundary  terms allows for a thermodynamically consistent description of the system. In Sec. 4, we implement the MEP, we identify the four phases that characterize the system, and we analyze   phase transitions and other  thermodynamic properties. In Sec. 5, we summarize and discuss our results. The Appendix contains a proof that the entropy of radiation in the box has no global maxima.

\section{Spacetime geometry for a shell of self-gravitating radiation}

\subsection{Constitutive equations}
The system under study is a spherical shell of self-gravitating radiation in thermal equilibrium. The thermodynamic equations for radiation are

\begin{equation}\label{eos}
\rho = 3P = b T^4, \qquad s=\frac{4}{3}b^{1/4}\rho^{3/4}
\end{equation}
where $\rho$ is the energy density, $P$ is the pressure, $T$ is the temperature, $s$ is the
entropy density and $b=\pi^2/15$ in  geometrized units  ($c=G=\hbar=k_B=1$).

The spacetime metric is static and spherically symmetric,
\begin{equation}\label{metric}
ds^2 = -L(r)^2 dt^2  + \frac{dr^2}{1-\frac{2m(r)}{r}} + r^2 (d\theta^2 + \sin^2\theta d\phi^2)
\end{equation}
where $L(r)$ is the lapse function, $m(r)$ is the mass function and $(t,r,\theta,\phi)$ is the usual coordinate system.

The radiation is contained between two reflecting spherical boundaries, an external boundary  at $r = R$, and an internal one at $r = r_0  < R$.

For $r> R$, the solution is Schwarzschild with ADM mass $M$, i.e.,
\begin{equation}\label{outmetric}
L(r ) =  \sqrt{1-\frac{2M}{r}}, \qquad\qquad m(r ) = M.
\end{equation}

For $r \in [r_0, R]$, the geometry is determined by the TOV equation for radiation,
\begin{equation}\label{struct1}
\frac{dm}{dr}=4\pi r^2\rho, \qquad \qquad \frac{d\rho}{dr} = -\frac{4\rho}{r^2}\frac{(m+\frac{4}{3}\pi r^3 \rho)}{1-\frac{2m}{r}}.
\end{equation}
For fixed ADM mass $M$, we solve the TOV equation from the boundary $r = R$ inwards. To this end, we must specify the density $\rho_R$ at $R$ (or equivalently the temperature $T_R$ at $r = R$), in addition to the mass $m(R)=M$.

The temperature at the boundary $T_R$ is related to the temperature at infinity $T_{\infty}$ by Tolman's law, $LT = T_{\infty}$,
\begin{equation}\label{Tinf}
T_\infty = T_R \sqrt{1-\frac{2M}{R}}.
\end{equation}

Tolman's law also determines the lapse function for $r \in [r_0, R]$,
\begin{equation}\label{medmetric}
L(r) = \sqrt{1-\frac{2M}{R}}\bigg(\frac{\rho_R}{\rho(r)}\bigg)^{1/4}
\end{equation}

Let $m_0 := m(r_0)$ and $\rho_0 := \rho(r_0)$. The solution for $r < r_0$ is Schwarzschild, since there is no matter present. In particular,
\begin{equation}\label{inmetric}
L(r) = \kappa \sqrt{1-\frac{2m_0}{r}}  \qquad \qquad m(r) = m_0,
\end{equation}
where $\kappa$ is a scaling constant for the  time coordinate $t$. By continuity at $r = r_0$,
\begin{equation}\label{kappa}
\kappa =  \sqrt{\frac{1- 2M/R}{1- 2m_0/r_0}}\bigg(\frac{\rho_R}{\rho_0}\bigg)^{1/4}
\end{equation}

If $m_0 > 0$, there is a Schwarzschild horizon at $r = 2m_0$
\footnote{The integration of the TOV equations from the boundary inwards never encounters a horizon, so $2m_{0} < r_{0}$ always.}
. If $m_0 = 0$, the geometry for $r < r_0$ is Minkowskian. If $m_0 <0$, the geometry for $r < r_0$ is Schwarzschild with
negative mass, i.e., there is a naked curvature singularity at the center.

\subsection{Solution curves}\label{solcurves}
Here, we consider the solution curves obtained when solving the TOV equations from the boundary inwards.

A solution of Eqs. (\ref{struct1}) is uniquely specified by the boundary data $(R, M, T_R)$.
It is convenient to introduce the variables,
\begin{equation}\label{uvdef}
\xi := \log\frac{r}{R} \qquad u:=\frac{2m(r)}{r} \qquad v:=4\pi r^2 \rho(r),
\end{equation}
which allow us to write the TOV equations as
\begin{equation}\label{struct2}
\frac{du}{d\xi} = 2v-u, \qquad \qquad \frac{dv}{d\xi}= \frac{2v(1-2u-\frac{2}{3}v)}{1-u}.
\end{equation}

Eqs. (\ref{struct2}) are simpler than Eqs. (\ref{struct1}), because they define an autonomous two-dimensional  dynamical system.
They need only two  positive numbers $(u_R, v_R)$ for boundary data, the outer boundary corresponding to $\xi = 0$. The center corresponds to $\xi \rightarrow - \infty$.

This simplification is possible only in linear equations of state, because they introduce no scale into the TOV equations. Hence, the solutions are invariant under the transformation $r \rightarrow \lambda r, m \rightarrow \lambda m $, and $\rho \rightarrow \lambda^{-2}\rho $, for any $\lambda > 0$.

\begin{figure}[H]
\centering
{{\includegraphics[width=12cm]{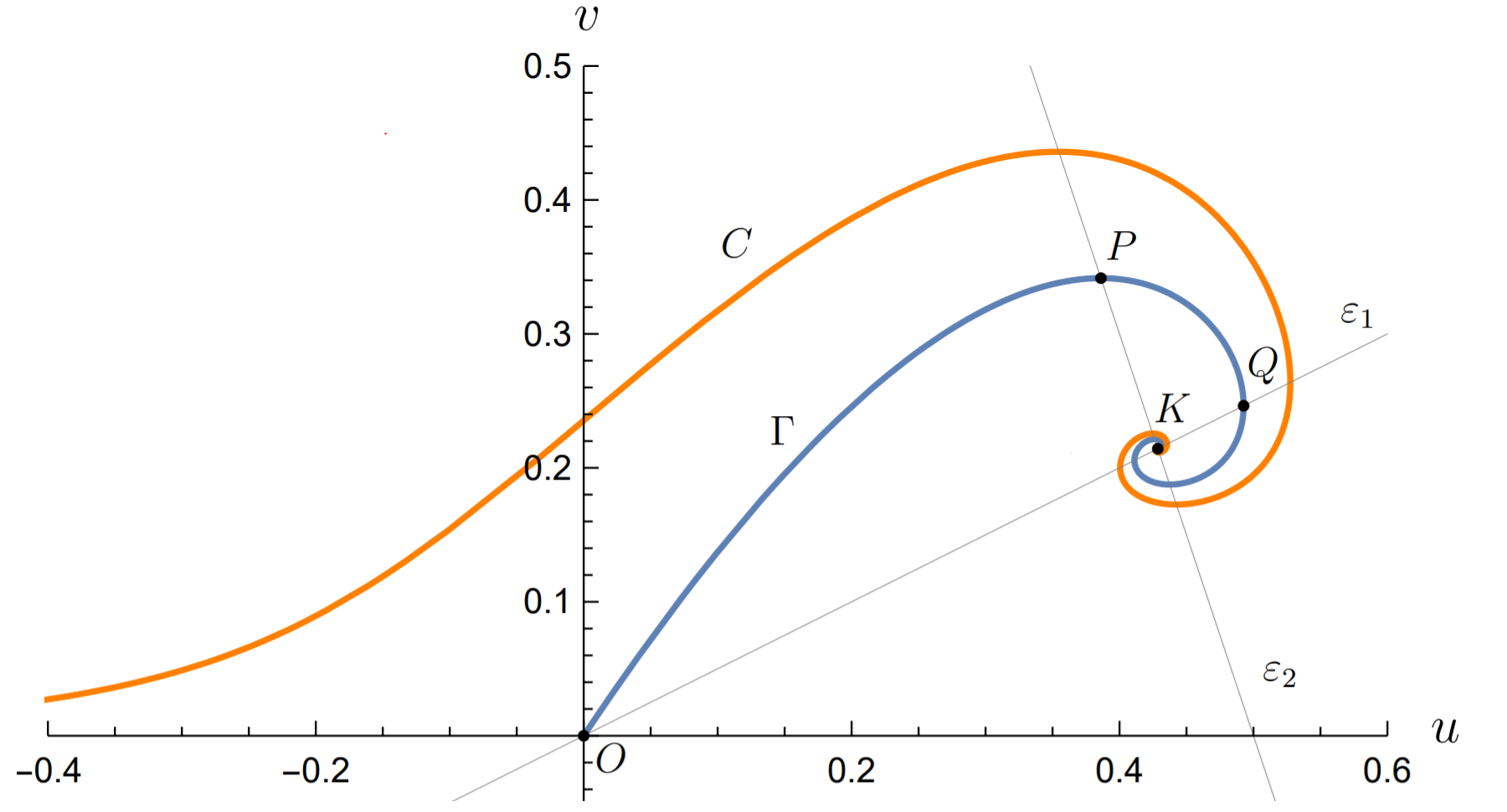} }}
\caption{Solution curves to the Tolman-Oppenheimer-Volkoff equation for radiation.}
\label{fig:curves}
\end{figure}

In order to describe the behavior of the solution curves, we identify two straight lines on the $u-v$ plane (Fig.\ref{fig:curves}):
\begin{itemize}
    \item $\varepsilon_1: 2v-u=0$, at which all solution curves satisfy $du/d\xi = 0$.
    \item $\varepsilon_2: 1-2u-\frac{2}{3}v=0$, at which all solution curves satisfy $dv/d\xi = 0$.
\end{itemize}
The two curves intersect at the point $K = (\frac{3}{7}, \frac{3}{14})$. $K$ and $O = (0, 0)$ are the equilibrium points of the dynamical system (\ref{struct2}).

A typical solution curve $C: u=u(\xi),\, v=v(\xi)$ satisfies the following properties \cite{AnSav21}.

\begin{enumerate}[(i)]
    \item There is a point $r_1 <R$, such that $u(r) = 0$. Furthermore, $u(r) < 0$ for all $r < r_1$.
    \item There is a point $r_2 < r_1$, such that $\frac{dP}{dr}(r_2) = 0$. Furthermore, $\frac{dP}{dr}> 0$, for $r < r_2$.
    \item $\lim_{\xi \rightarrow - \infty} (u,v) =  (-\infty, 0) $,
   \item  $\lim_{\xi \rightarrow  \infty}  (u,v) = K  $.
\end{enumerate}

The only non-trivial exception is the curve $\Gamma$ of regular solutions, i.e., solutions that satisfy $m(0)=0$. On $\Gamma$, $\lim_{\xi \rightarrow - \infty} (u,v) =  O$. The variables $v$ and $u$ attain  maximum values on $\Gamma$ at the points $P$ and $Q$ respectively, where   $(u_P, v_P) \approx (0.3861, 0.3416)$ and $(u_Q, v_Q) \approx (0.4926, 0.2463)$.

The following solution curves are degenerate cases. (i) The points $K$ and $O = (0, 0)$ are equilibrium points, so each defines a distinct solution curve. (ii) A point on the $u$ axis  ($v_R = 0, u_R \neq 0$), evolves with   $u(\xi) = u_R e^{-\xi},\, v(\xi) = 0$, and it encounters an even horizon. This corresponds to a Schwarzschild black hole without matter.

\subsection{Shell configurations}
The solution curves uniquely determine the spacetime geometry associated to a shell of radiation---for a past treatment of this system, see, Ref. \cite{Kim17b}. We select $R, M, T_R$, and we follow the solution curve until we encounter $r_0$. The segment of the solution curve between $r_0$ and $R$ determines the shell's geometry, for $r > R$, and $r < r_0$ the geometry is Schwarzschild with mass $M$ and $m_0$ respectively.

Hence, the space $Z$ of equilibrium configurations for a shell of self-gravitating radiation is four-dimensional. It can be described by the coordinates  $(R, r_0, M, T_R)$ or
equivalently by the coordinates
$(R, \xi_0, u_R, v_R)$, where $\xi_0 = \log(r_0/R)$.

The solutions classes fall into three types, depending on the value of $m_0 = m(r_0)$.
\begin{itemize}
\item $m_0 = 0$: Type F (flat) solution for $r < r_0$.
\item $m_0 > 0$: Type B (black-hole), it contains a Schwarzschild horizon at $r < r_0$.
\item $m_0 < 0$: Type S (naked singularity), it contains the negative-mass Schwarzschild singularity $r < r_0$.
\end{itemize}

 F-type solutions form a set of measure zero in $Z$ that acts as a the boundary between the subset of B-type and S-type solutions. The curve of the F-type solutions on the $u_R-v_R$ plane (the {\em F-curve} for brevity ) depends only on   $\xi_0$, because of the scaling symmetry.

\begin{figure}[H]
    \centering
   {{\includegraphics[width=14cm]{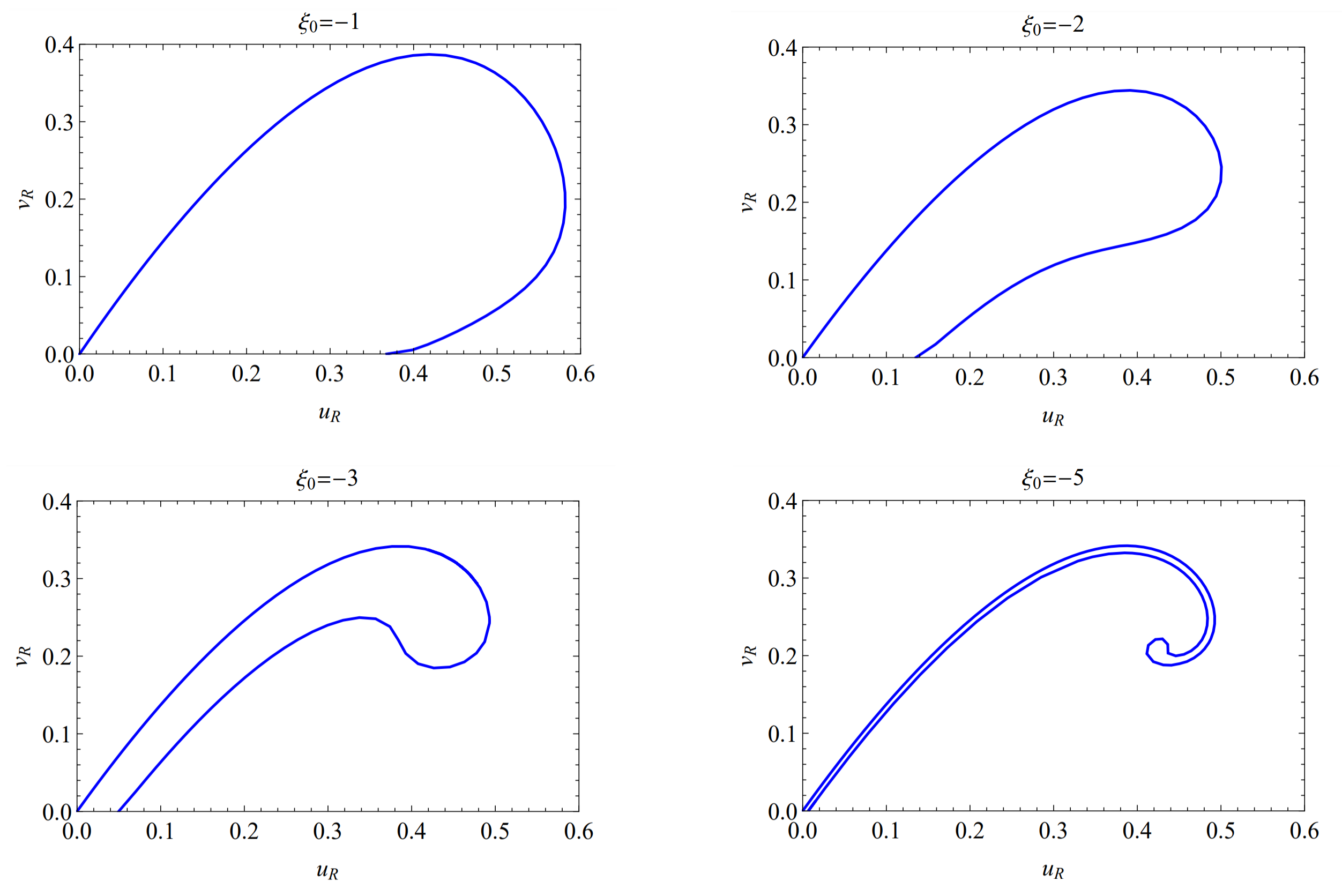} }}
  \caption{The F-curve for different values of $\xi_0$, namely. Type B configurations lie in the region between the R-curve and the horizontal axis. Outside this region there exist only Type S configurations. As $\xi_0$ decreases the F-curve shrinks to the solution curve $\Gamma$.}
  \label{fig:typeRcurve}
\end{figure}

 In Fig. \ref{fig:typeRcurve}, the F-curve is plotted for different  $\xi_0$. B-type solutions lie in the region between the F-curve and the $u_R$ axis; the remainder corresponds to S-type solutions. As $\xi_0$ decreases, so does the area of the $B$ phase. At $\xi \rightarrow - \infty$,  B-type solutions disappear and the F-curve coincides with the line $\Gamma$ of Fig.1.

For fixed $\xi_0$, the F-curve has two distinctive points.
\begin{itemize}
    \item The point $u_{max}(\xi_0)$ that corresponds to the maximum value of $u_R$.
    \item The final point $u_f(\xi_0) \neq 0$, where the curve intersects the horizontal axis ($v_R = 0$).
\end{itemize}

\begin{figure}[H]
    \centering
   {{\includegraphics[width=8cm]{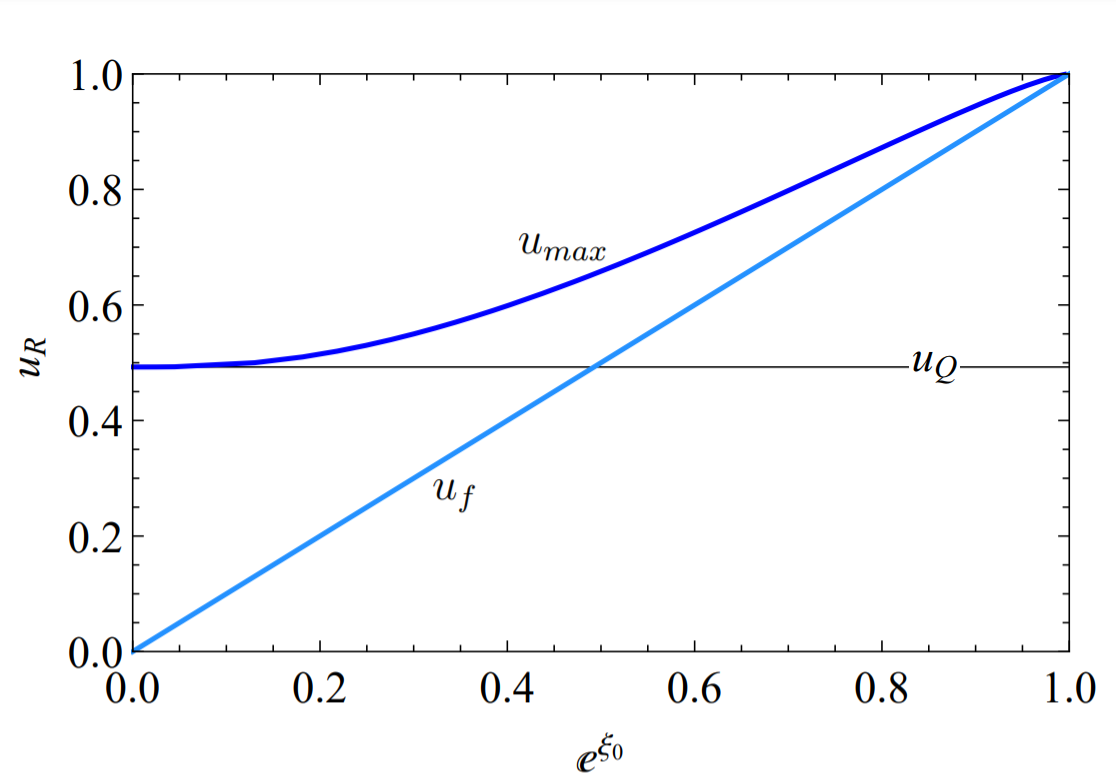} }}
  \caption{Plot of $u_{max}$ against $\xi_0$. Notice that $\lim_{\xi_0\to -\infty}u_{max} = u_Q$. }
  \label{fig:uOV}
\end{figure}

From Fig.\ref{fig:typeRcurve}, we see that  both $u_{max}(\xi_0)$ and $u_f(\xi_0)$ decrease with decreasing $\xi_0$---see, Fig. (\ref{fig:uOV}). As $\xi_0 \rightarrow -\infty$, $u_{max} \rightarrow u_Q$, i.e., it corresponds to the Oppenheimer-Volkoff limit for a sphere of radiation \cite{PWZ}. In the same limit, $u_{f}$ vanishes as $e^{-\xi_0}$. The dependence of $u_{max}$ and of $u_f$ on $\xi_0$ is plotted in Fig.  (\ref{fig:uOV}).

From Fig. \ref{fig:typeRcurve}, we can analyse how the three types of solution are distributed in the one-dimensional submanifolds (fibers) $V_{(u_R, \xi_0, R)}$ of constant $(u_R, \xi_0, R)$. For  given $\xi_0$, each  fiber corresponds to a line $u_R =$constant in the plots of  \ref{fig:typeRcurve}. We characterize the fibers in terms of their intersection with the R-curve. There are three types of behavior, by which  we characterize the fibers as being of type I, II, and III.

\begin{itemize}
    \item Fibers of type I are defined by $u_R \leq u_f(\xi_0)$. The  line $u_R =$ constant  intersects the F-curve only once, at some point $v_1$. For $v_R < v_1$, the fiber contains solutions of type B, and for $v_R> v_1$, it contains solutions of type S.
    \item Fibers of type II are defined by $u_f(\xi_0) < u_R < u_{\max}(\xi_0)$.
    The  line $u_R =$ constant  intersects the F-curve at least twice. Hence, these fibers involve
 two solutions of type F, at $v_R = v_1$ and $v_R = v_2$, such that all solutions with $v \notin [v_1, v_2]$ are of type S. In the interval $(v_1, v_2)$ the solutions are either of type B, or there exist alternating regions of type B and type S solutions. The latter is the case for large negative values of $\xi_0$ where the R-curve develops a spiraling shape, hence, it intersects the line $u_R =$ constant more than twice.

    \item Fibers of type III are defined by $u_R > u_{\max}(\xi_0)$. All points in these fibers correspond to solutions of type S.

\end{itemize}


 \section{Thermodynamic consistency and Maximum Entropy principle}

\subsection{The fundamental representation}
We proceed with a study of the thermodynamics of the shell system. We will be working with the fundamental representation (i.e., the entropy representation) of thermodynamic systems, in which the relevant thermodynamic potential is the entropy $S$, expressed as a function of the total energy $M$ and the constraints in the extension of the system, namely the area of the bounding surfaces, as expressed in the variables $r_0$ and $R$.

In extensive systems,  the choice of representation does not affect physical predictions. Extensivity together with the second law of thermodynamics imply that $S$ is a concave function of the extensive variables \cite{Callen}. The Legendre transform for concave functions fully preserves their information: it is an involution, i.e., the double Legendre transform of a functions $S$ returns the original function $S$.
Since the different representations are connected by a Legendre transform, they all contain the same information about the system.

In a non-extensive system, the entropy $S$ needs not be a concave function. Typically,  {\em  convex insertions appear}, i.e., regions of the fundamental thermodynamic space where  $S$ is convex. The double application of the Legendre transform  does not return the original function $S$ but its concave hull \cite{inequiv}\footnote{The Maxwell construction employed in the study of first-order phase transitions is a well known example of taking the concave hull of a non-concave entropy function. However, the Maxwell construction is physically meaningful only in extensive systems, where concavity of the entropy is a physical necessity.}. In this sense, the Legendre transform does not preserve all thermodynamically relevant information. As a result, the other thermodynamic representations are not equivalent to the fundamental one.

At the microscopic level, this difference is manifested in the inequivalence between the microcanonical and the canonical ensembles. The equivalence of these ensembles in ordinary statistical mechanics follows from the requirement  that the size of the system is much larger than the range of the force between the constituents. Then, the entropy obtained from the microcanonical distribution is concave \cite{Ruelle}. In presence of long-range forces, the microcanonical entropy is generically non-concave, leading to the inequivalence of ensembles. A convex insertion in some region of the fundamental space implies the presence of a  first-order phase transition \cite{Touch, BouBa}.

Hence, when studying an isolated gravitational system, it is necessary to use the fundamental representation, as any other representation will misrepresent the physics\footnote{In Ref. \cite{AnSav14}, it was shown that the natural representation for a self-gravitating system is a {\em free entropy} representation, based upon the thermodynamic potential $\Omega$ (the free entropy)  that is obtained as a Legendre transform of the entropy with respect to the number $N$ of particles in a system. For radiation, the number of particles is not conserved, and $\Omega$ coincides with the entropy function.}. Typically, the other representations (Gibbs, Helmholtz, enthalpy) are obtained by coupling the system to an external reservoir. We mentioned previously that   idealized couplings to a reservoir make no sense in a gravitating system, because the reservoir can affect even the constitutive equations of the system.

The shell of radiation considered here can be viewed as a gravitating reservoir for the system in the interior region. However, radiation is strongly affected and it strongly affects the interior region, so that a split between a system and a reservoir makes no sense.   In gravitational systems, {\em we must treat system and reservoir as a single isolated system}, and for this reason, we must always work with the fundamental representation.

\subsection{The maximum-entropy principle}

First, we consider the shell system with gravity switched off. In the entropy representation, the entropy $S$ is a function of $R, r_0$ and $M$, \begin{eqnarray}
S(R, r_0, M) = \frac{4}{3}b^{1/4} V^{1/4}(R, r_0) M^{3/4},
 \end{eqnarray}
 where $V(R, r_0) = \frac{4\pi}{3}(R^3-r_0^3)$ is the volume of the shell. The thermodynamic state space $Q = \{(R, r_0, M)\}$ is three-dimensional.

 The thermodynamic state space remains the same when gravity is switched on.   Since the volume is a variable in a gravitational system, the dependence of $S$ on $R$ and $r_0$ is non trivial. The problem now is that the set $Z$ of solutions to the TOV equation is four dimensional. Hence, the independent thermodynamic variables do not fix uniquely the solution.

In compact stars ($r_0=0$), this problem is usually addressed by an additional assumption, namely, regularity at the center $m(0)=0$. Regular solutions form a set of measure zero in the set of all solutions. Almost all solutions have $m(0)<0$, and there are no  solutions with $m(0) > 0$. Solutions with $m(0) < 0$ have a curvature singularity at the center. However, this singularity causes no problems with causality and predictability: the spacetime has no inextensible geodesics, it is bounded-acceleration complete, and it is conformal to a globally hyperbolic spacetime with boundary \cite{AnSav21}.

The problem with the  regularity condition is that it does not cover the whole thermodynamic state space, as there are no solutions with $m(0) = 0$, if $M> M_{OV}$, where $M_{OV}$ is the Oppenheimer-Volkoff limit. In Ref. \cite{AnSav12}, it was also argued that there is no good reason to {\em a priori} exclude singular solutions from all considerations, since they would appear in the sum of geometries of a quantum theory of gravity, at least as virtual solutions.

 An important benefit of using a shell as our thermodynamic system is that it demonstrates unambiguously the inadequacy of the regularity condition for selecting equilibrium configurations. Even if one wants to {\em a priori} exclude solutions of type S---presumably because they involve a naked singularity---, there is no justification in excluding solutions of type B. Hence, the regularity condition cannot identify an equilibrium configuration for the radiation shell.

 We will select the equilibrium configurations by employing the {\em maximum entropy principle}. The MEP asserts that the values assumed by any parameters in absence of constraints are those that maximize the entropy over the manifold of constrained states \cite{Callen}.

 In the entropy representation,  the total mass $M$ and the shell radii $R$ and $r_0$ are assumed to be constrained. As shown in Sec. 2.3, the manifold $Z$ is foliated by surfaces of constant $(M, R, r_0)$, or equivalently, of  constant $(R, u_R, \xi_0) $. Each fiber $V_{(R, u_R, \xi_0)}$ of the foliation is parameterized by $v_R$.

 We can construct an entropy function $S_Z$ on $Z$, $S_Z(R, u_R, \xi_0, v_R)$.
 The MEP asserts that the equilibrium state for  $M, R, r_0$ is obtained by maximizing  the entropy functional along the associated fiber,
 \begin{eqnarray}
 S_{eq}(M, R, r_0) = \mbox{max}_{v_R} S_Z(R, u_R, \xi_0, v_R). \label{Sz}
 \end{eqnarray}
 If there is  no global maximum of $S_Z$ on a fiber, the MEP fails to apply. This is the case in some gravitating systems, known as the {\em gravothermal catastrophe} \cite{gravothermal}.

 In what follows, we will show that if
$S_Z(R, u_R, \xi_0, v_R)$ involves only a contribution $ S_{rad} $ from radiation, the shell system is not thermodynamically consistent. In contrast, an appropriate gravity contribution $S_{gr}$ makes the system consistent. The gravitational entropy $S_{gr}$ is a Noether charge associated to the spacetime boundaries in the region $r < r_0$. For  B-type solutions, $S_{gr}$ is the Bekenstein-Hawking entropy. For the S-type solutions, the working expression for $S_{gr}$ is the same with the one identified in Ref. \cite{AnSav12}

From a thermodynamic point of view, the need of a term $S_{gr}$ from $r < r_0$ is obvious. If we view the radiation shell as a thermal reservoir in contact with the interior region, then it is necessary to include a contribution from the interior region, otherwise a black hole would be thermodynamically indistinguishable from flat space. Since the interior solution is vacuum, the only contribution to entropy can be of gravitational origin.

 \subsection{Radiation entropy}

The radiation entropy $S_{rad}$ of a solution to the TOV equation is given by
\begin{eqnarray}
S_{rad} = 4 \pi \int_{r_0}^R \frac{dr r^2 s}{\sqrt{1- \frac{2m(r)}{r}}}
= \frac{4}{3} (4\pi b)^{1/4} \int_{r_0}^R dr \frac{r^{1/2} v^{3/4}}{\sqrt{1-u}}.
\end{eqnarray}

Solutions to the TOV equation satisfy \cite{PWZ}
\begin{eqnarray} \label{dsdr}
 \frac{r^{1/2}v^{3/4}}{\sqrt{1-u}} = \frac{d}{dr}\bigg(  \frac{v + \frac{3}{2}u}{6 v^{1/4}\sqrt{1-u}} r^{3/2} \bigg),
\end{eqnarray}
from which we obtain
\begin{equation}
S_{rad}(R,\xi_0,u_R,v_R) = \frac{2}{9}(4\pi b )^{1/4}\left(
\frac{v_R+\frac{3}{2}u_R}{v_R^{1/4}\sqrt{1-u_R}} -
\frac{v_0+\frac{3}{2}u_0}{v_0^{1/4}\sqrt{1-u_0}}e^{3\xi_0/2}  \right)R^{3/2}, \label{srad0}
\end{equation}
where $u_0=u(\xi_0)$ and $v_0 = v(\xi_0)$.
The simple dependence of $S_{rad}$ on $R$ is due to the scaling symmetry.

\begin{figure}[H]
    \centering
   {{\includegraphics[width=10cm]{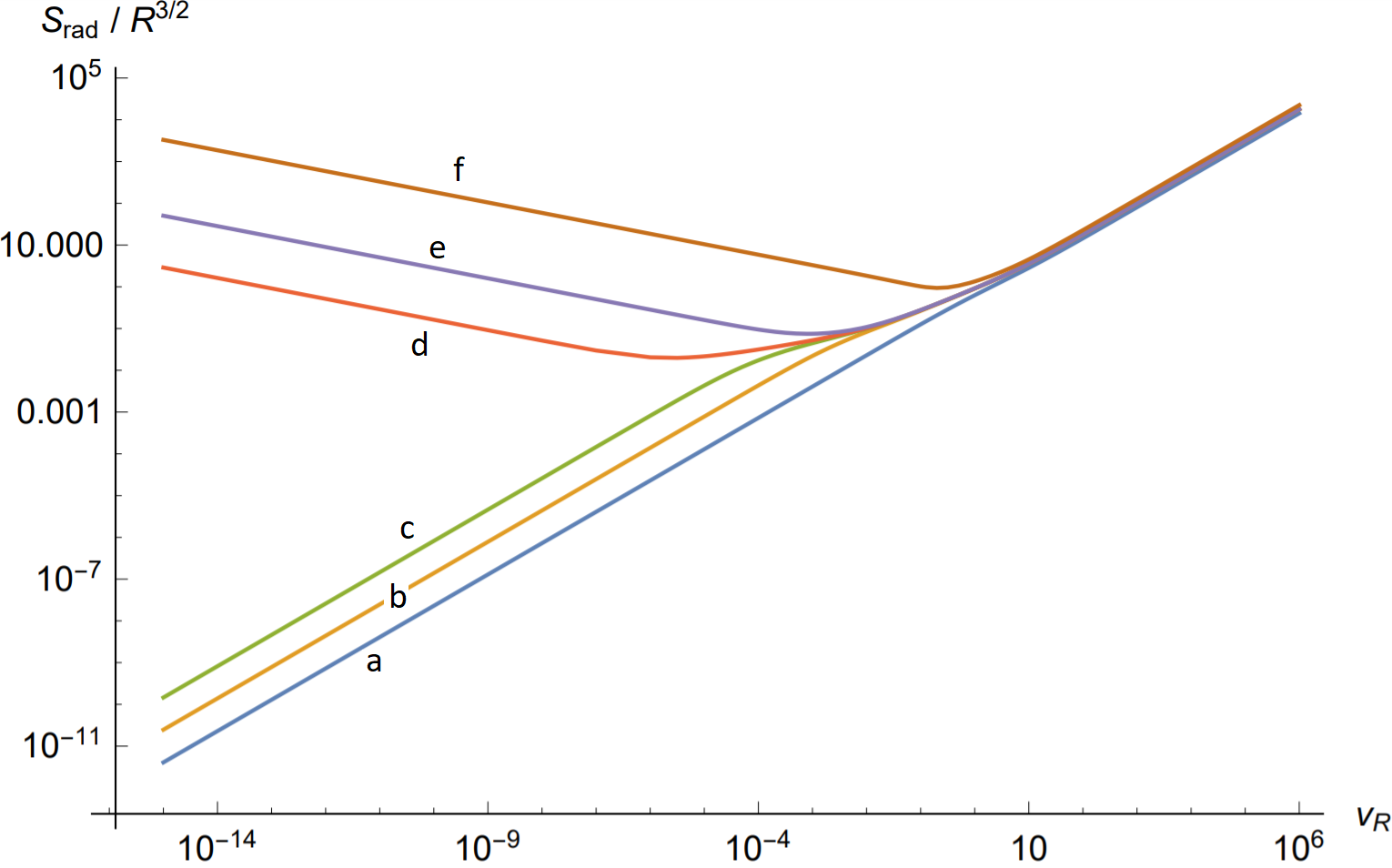} }}
  \caption{ $S_{rad}/R^{3/2}$ as a function of $v_R$ for fixed $\xi_0 = -1$ and for different values of $u_R$.(a) $u_R=0.01$ (b) $u_R=0.36$ (c) $u_R=0.367$ (d) $u_R=0.368$ (e) $u_R=0.37$ (f) $u_R=0.6$.}
  \label{fig:blowup}
\end{figure}

The key point here is that $S_{rad}$ has no global maximum. The reason is that it diverges at infinity. In particular, we found that
\begin{itemize}
\item For all $R, \xi_0, u_R$, $\lim_{v_R \rightarrow \infty}S_{rad} =  \infty$.
\item For all $R$, and for $\xi_0 < \log u_R$, $\lim_{v_R \rightarrow \infty} S_{rad} = \infty$.
\end{itemize}
These limits can easily be seen in numerical calculation. Fig. \ref{fig:blowup} shows the results of such a calculation, $S_{rad}/R^{3/2}$ is plotted as function of $v_R$ for fixed $u_R$ and $\xi_0$.

We also obtained an analytic proof of the entropy limits above, which is detailed in the Appendix A.  The proof is rather long, as it requires an explicit construction of the  solution curves for $v_R >> 1$ and for $v_R << 1$. Still, it is instructive as the same methods can be employed in the study of other solutions to the TOV equation.

To conclude, the MEP cannot be applied in the shell system if $S_{rad}$ is the only contribution to the system's entropy.There is no global maximum of entropy in the fibers of constant $(R, u_R, \xi_0)$.

\subsection{Entropy from spacetime boundaries}
Ever since Bekenstein's and Hawking's work on black hole thermodynamics, we know that entropy can be meaningfully assigned to gravitational degrees of freedom.

Wald showed that black hole entropy can be expressed in terms of the Noether charge $Q(\xi)$ of spacetime diffeomorphisms \cite{Wald93}, as
\begin{eqnarray}
S = \frac{Q(\xi)}{T_{\infty}},
\end{eqnarray}

The Noether charge $Q(\xi)$ is defined in terms of the time-like Killing vector $\xi = \frac{\partial}{\partial t}$, normalized so that $\xi^{\mu}\xi_{\mu}=-1$ at infinity, and evaluated on the horizon, viewed as a   boundary of the surfaces of constant $t$:
\begin{eqnarray}
Q(\xi) = \frac{\lambda}{4 \pi} \oint_{\partial \Sigma} d\sigma_{\mu \nu} \nabla^{\mu} \xi^{\nu},
\end{eqnarray}
where $\lambda$ is an arbitrary multiplicative constant.

For positive-mass Schwarzschild spacetime, $Q(\xi) = 2 \lambda M$, when evaluated at the horizon. Since $T_{\infty} = 8 \pi M$, the Bekenstein-Hawking entropy $S_{BH} = 4\pi M^2$ is obtained for $\lambda = \frac{1}{4}$.

We will use the  Bekenstein-Hawking entropy for the entropy of the horizon that appears in solutions of type $B$: $S_{grav} = 4 \pi m_0^2$.

For type S solutions, the singularity at $r = 0$ defines a timelike boundary \cite{AnSav21}. For this boundary,
\begin{eqnarray}
Q(\xi) =  2 \lambda m_0\kappa,
\end{eqnarray}
suggesting an entropy associated to singularity equal to
\begin{eqnarray}
S_{sing} = \frac{2 \lambda m_0\kappa}{ T_{\infty}} =  \lambda (4\pi b)^{1/4} \frac{u_0}{v_0^{1/4}\sqrt{1-u_0}}e^{3\xi_0/2} R^{3/2}.
\end{eqnarray}

The only way we have found to specify $\lambda$ is through the requirement of thermodynamic consistency. In \cite{AnSav12}, it was shown that the only way to implement the MEP for a sphere of self-gravitating radiation is by assigning entropy to the singularity with
$\lambda = 2$.

\begin{figure}[H]
    \centering
   {{\includegraphics[width=10cm]{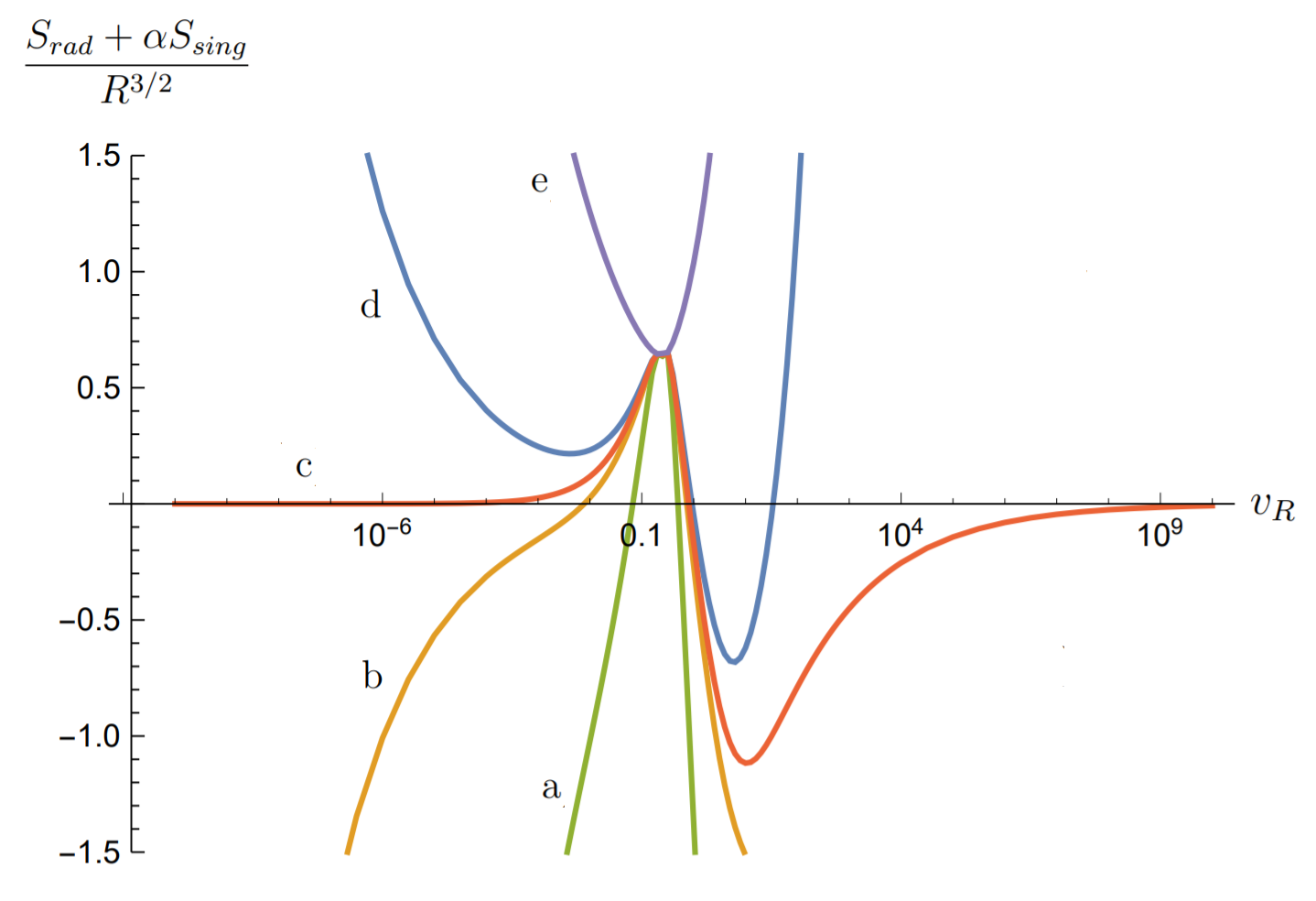} }}
  \caption{Plot of the candidate entropy function $(S_{rad}+\alpha\cdot S_{sing})R^{-3/2}$ against $v_R$ for $(u_R, \xi_0)=(0.44,-4)$ and different values of $\alpha$ (a)$\alpha=0 $, (b)$ \alpha=0.9$, (c) $\alpha=1$, (d) $\alpha=1.08$ and (e) $\alpha=2$. Any $\alpha < 1$, the entropy function is unbounded from below, and for $\alpha > 1$, it has not global maximum. Only the case $\alpha = 1$ corresponds to a physically admissible entropy function.}
  \label{fig:finite}
\end{figure}

The same holds in the shell system studied here. The only value of $\lambda$ that provides a consistent implementation of the MEP is $\lambda = 2$. This is best seen in Fig. (\ref{fig:finite}).There we define $S_{sing}$ for $\lambda = 2$, and consider candidate entropy functions $S_Z := S_{rad}+\alpha S_{sigma}$ or different values of $\alpha$. Only the value $\alpha = 1$ leads to a thermodynamically consistent function $S_Z$: for $\alpha < 1$, $S_Z$ is not bounded from below\footnote{An entropy function that is not bounded from below cannot be consistent with the statistical interpretation of entropy, where entropy is proportional to the logarithm of the number of microstates. If there is a lower bound, we can always add a constant to the entropy function and render it positive.}, and for $\alpha > 1$, $S_Z$ has no global maximum on fibers.

In what follows, we will take $S_{sing}$ with $\lambda = 2$ as the gravitational contribution to the total entropy in the S phase. $S_{sing}$ is negative, and vanishes for $m_0=0$, thereby enhancing the stability of $R$-type solutions\footnote{We have tested this method to systems described by different equations of state, (e.g., fermionic  matter) and we have found that the value $\lambda =2$ is the only one that gives a consistent MEP. At the moment, we lack a fundamental explanation of this fact. }.

\begin{figure}[H]
   {{\includegraphics[width=17cm]{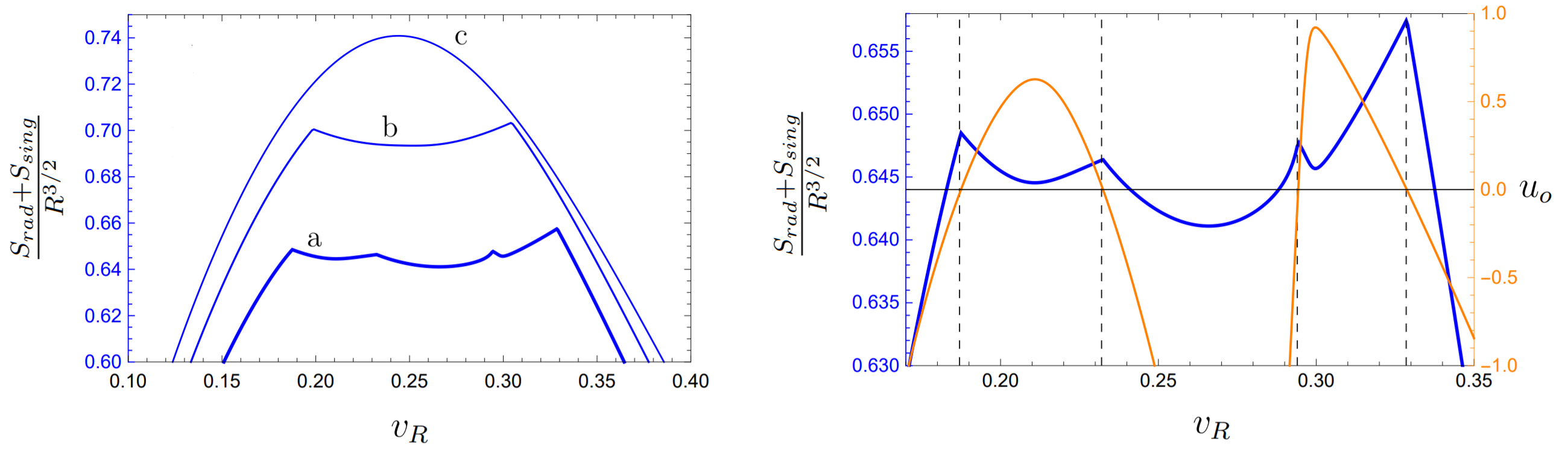} }}
  \caption{Left: We plot $\frac{S_{rad}+S_{sing}}{R^{3/2}}$ as a function of $v_R$ for $\xi_0=-4$ and for different values of $u_R$: (a) $u_R=0.44$  (b) $u_R=0.47$, and   (c) $u_R=0.50$. Case (a) has four F-type solutions, which correspond to the local maxima of the entropy function. Case (b) has two F-type solutions, again corresponding to local maxima of entropy.
  Case (c) has no F-type solution, the entropy maximum corresponds to a S-type solution.
  \\
   Right:  $\frac{S_{rad}+S_{sing}}{R^{3/2}}$ (blue) and $u_0$ (orange) are plotted as function of $v_R$ for $\xi_0=-4$ and $u_R=0.44$. This plot demonstrates clearly the one-to-one correspondence between maxima of entropy and F-type solutions ($u_0 = 0$).   }
  \label{fig:conemaxreg}
\end{figure}

We found numerically that {\em  all F-type solutions at a fiber of constant $(u_R, R, \xi_0)$ correspond to local maxima of $S_{rad}+S_{sing}$}, with respect to $v_R$. S-type maxima are only possible in fibers with no F-type solutions, i.e., for $u_R > u_c(\xi_0)$. This behavior is demonstrated in Fig. \ref{fig:conemaxreg}. This result is  identical with the one of Ref. \cite{AnSav12} for balls of radiation, and it also persists for other equations of state.

We conclude that the entropy function $S_Z(R, \xi_0, u_R, v_R)$ is
\begin{eqnarray}
S_Z = \left\{ \begin{array}{cc} S_{rad} + S_{BH}, & u_0 \geq 0 \\ S_{rad} +S_{sing},& u_0 < 0 \end{array} \right..
\end{eqnarray}
We note that both $S_{rad}(R, \xi_0, u_R, v_R) $ and $S_{sing}(R, \xi_0, u_R, v_R) $ can be expressed as $f(\xi_0, u_R, v_R)R^{3/2}$, i.e., they scale with $R^{3/2}$. In contrast,
 \begin{equation}
    S_{BH} = 4\pi m_0^2 = \pi (u_0 e^{\xi_0})^2 R^2
\end{equation}
scales with $R^2$. The black hole contribution breaks the scaling invariance of the entropy that originates from the scale independence of the equation of state.

\section{Phase transitions and other thermodynamic properties}

\subsection{The four phases of the system}

Next, we implement the MEP on each fiber of constant $(R, u_R, \xi_0)$. There are three different scenarios, one for each type of fiber, see, Sec. 2.3.

\medskip

\noindent {\em Type I fibers:} If $u_R < u_f(\xi_0)$, then the fiber contains a single solution of type F, say at $v_R = v_1$, that is a local maximum. Solutions for $v_R > v_1$ are of type S, and they have all smaller entropy than the F-type solution. There is no local maximum of entropy for S-type solutions.

For $v_R < v_1$, solutions are of type B.  The maximum value of entropy for B-type solutions occurs typically at very small $v_R$, often numerically indistinguishable from  $v_R = 0$, i.e., a black hole with very little radiation in the shell.  We will refer to this type of black hole, as a solution of type $B_I$.

Hence, the entropy along a type I fiber contains one local entropy maximum $S_F$ of type F, and one local maximum $S_{B_I}$ of type $B_I$.  The global maximum is the highest of the two local maxima.

\begin{figure}[H]
    \centering
   {{\includegraphics[width=13cm]{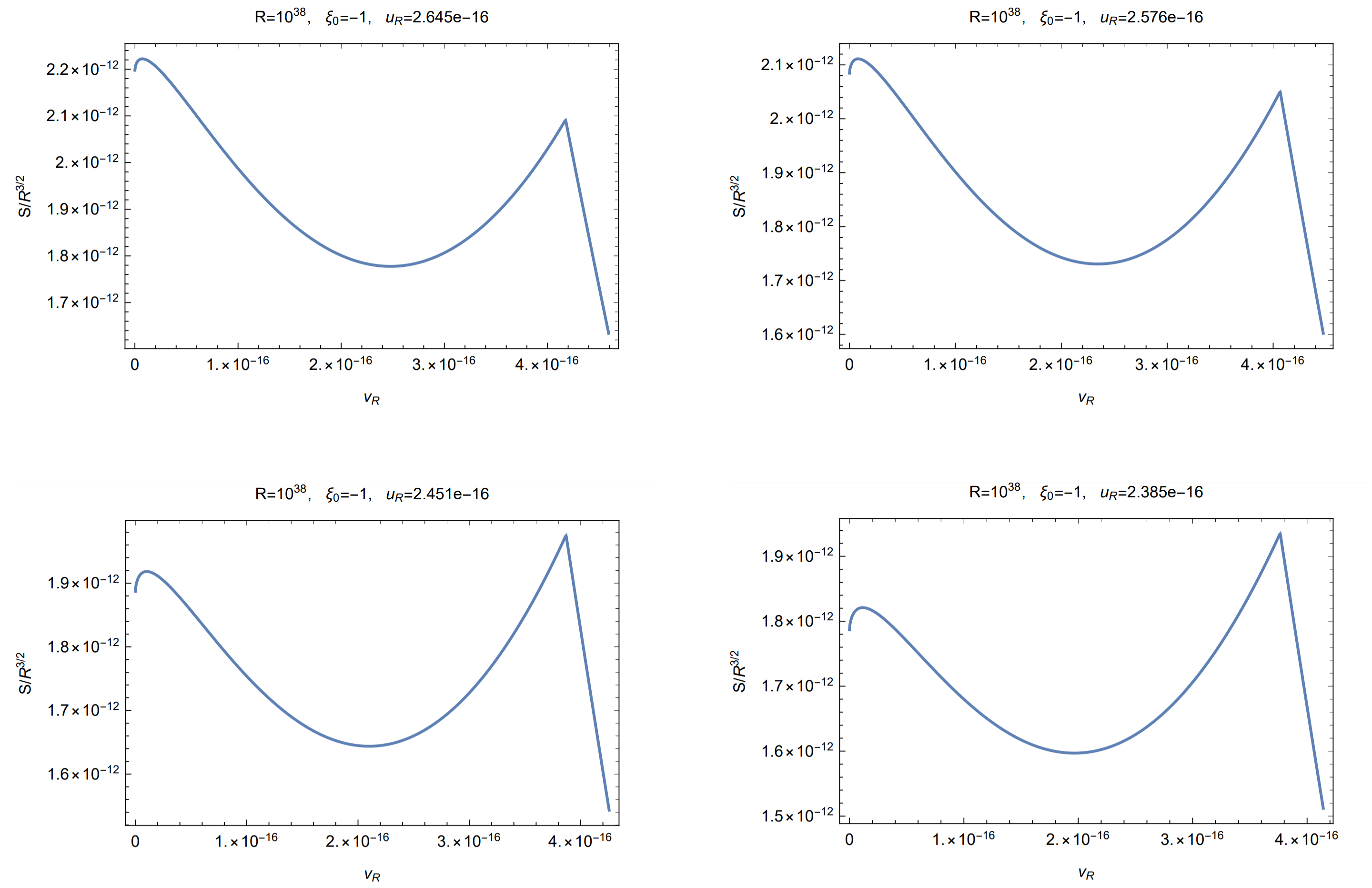} }}
  \caption{Entropy maxima and phase transition in fibers of type $I$.}
  \label{fig:BRlab}
\end{figure}

Typical plots of the entropy along  a fiber are given in Fig. \ref{fig:BRlab}. The behavior is characteristic of a first-order phase transition. As the location of the fiber changes, so does the relative height of the two maxima.
If $S_{B_I} > S_R$, the equilibrium phase is of type $B_I$; if $S_F > S_{B_I}$ the equilibrium phase is of type F. The submanifold of $Q$ where  $S_{B_I} = S_F$ is the coexistence curve for the B-F transition.

\medskip

\begin{figure}[H]
    \centering
   {{\includegraphics[width=13cm]{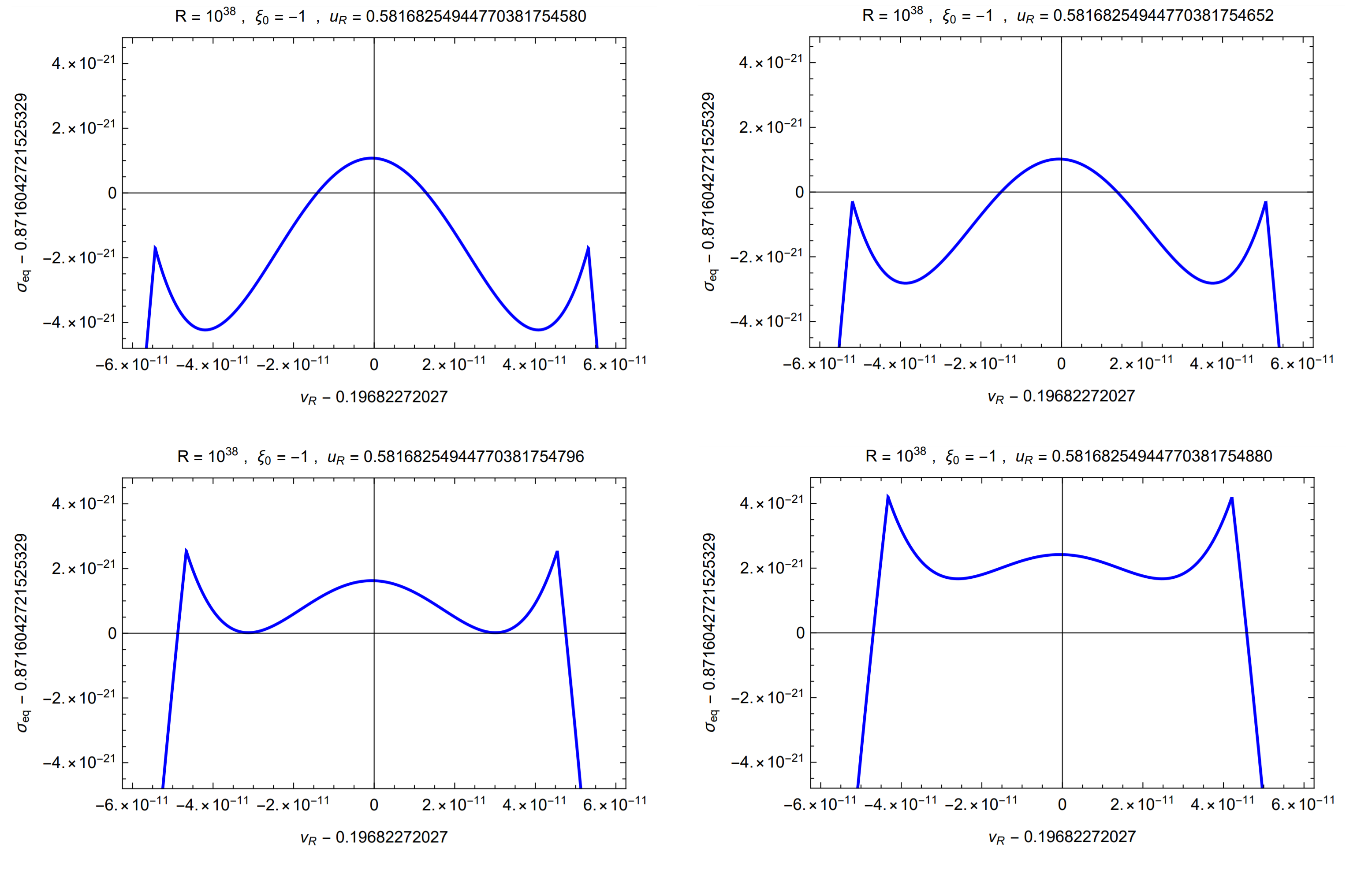} }}
 \caption{Entropy maxima and phase transition in fibers of type $II$ }
  \label{fig:BROV}
\end{figure}

\noindent {\em Type II fibers:} Type II fibers are characterized by several local maxima of  type F and at least one local maximum of  type B. The main difference from fiber I is that the B-type local maxima lie at an intermediate value between  two F-type maxima---see, Fig. \ref{fig:BROV}.

In these B-type solutions a significant fraction of mass is in the form of radiation in the shell, and the black hole horizon is often very small.
We will denote these solutions as $B_{II}$t

\medskip

\noindent {\em Type III fibers:}
Type III fibers contain only S type solutions. Obviously, the entropy-maximizing solutions are of type S.

\medskip

To summarize, the implementation of the MEP demonstrates that the thermodynamic state space $Q$ splits into four components, which corresponds to phases of types $R$, $B_I$, $B_{II}$ and $S$. The results of this analysis are summarized in Table \ref{table:1}.

\begin{table}[h!]
\centering
\begin{tabular}{ |c|c|c|c| }
 \hline
  Fiber type & Definition & Types of solution & Phases  \\
 \hline\hline
 I & $u_R < u_f(\xi_0)$ & S, F, B & F, $B_I$ \\
 \hline
 II& $u_f(\xi_0) < u_R < u_{max}(\xi_0)$ & S, F, B &  F,$ B_{II}$ \\
 \hline
 III & $u_R >u_{max} (\xi_0)$ & S &   S  \\
 \hline
\end{tabular}
\caption{The three types of fiber and the four phases.}
\label{table:1}
\end{table}

\subsection{Phase diagrams and coexistence curves}

\begin{figure}[H]
    \centering
   {{\includegraphics[width=14cm]{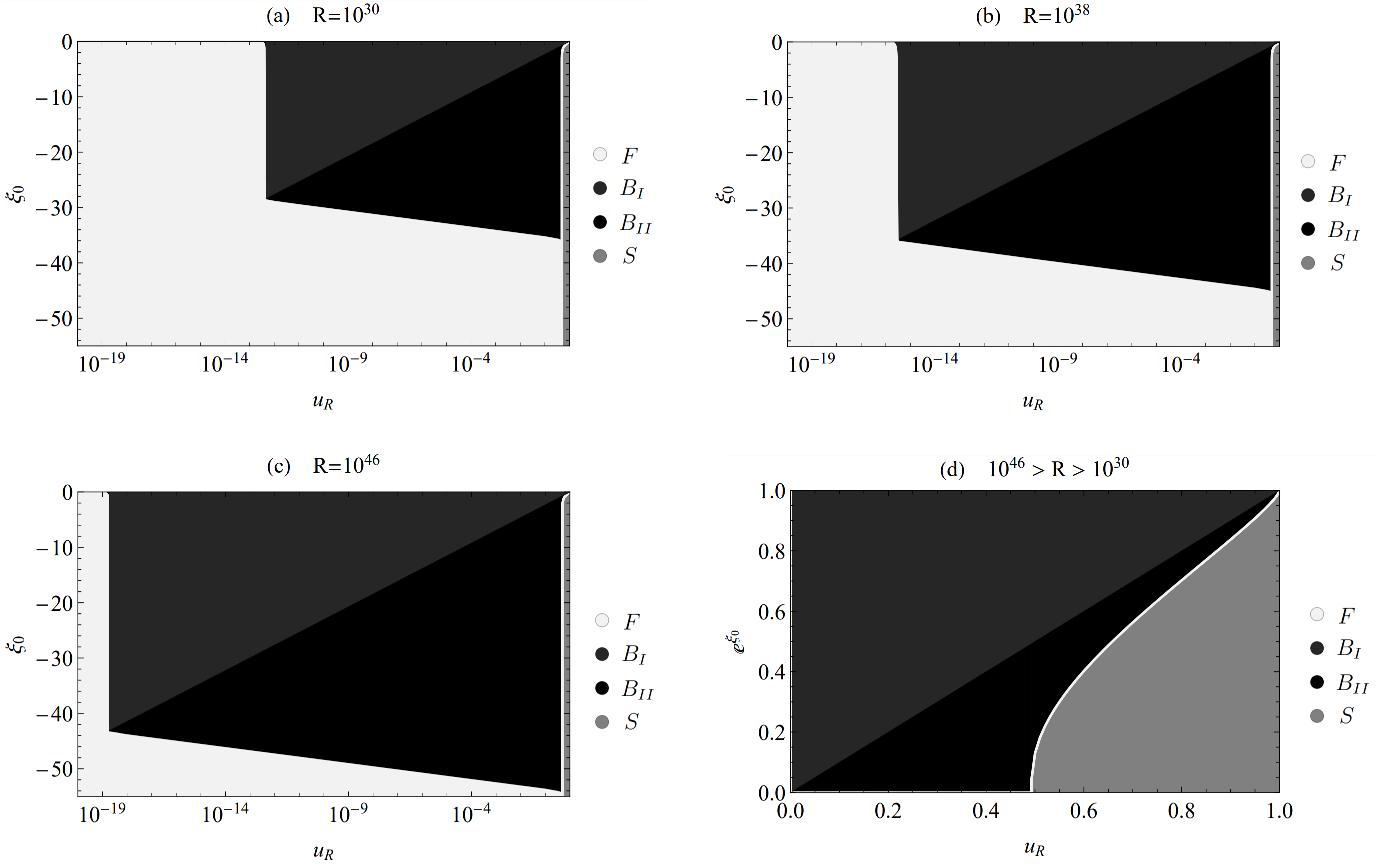} }}
  \caption{Phase diagrams for different values of $R$. In plots (a), (b) and (c) we use a logarithmic scale for the $u_R$ axis. In plot (d), we use a linear scale for the $u_R$ axis, the plot is practically insensitive to $R$, since the differences at small $u_R$ cannot be distinguished. }
  \label{fig:mepphd}
\end{figure}

In Fig. (\ref{fig:mepphd}), we show how the submanifolds of constant $R$ are partitioned into the four phases, for different values of $R$.

We remark the following.

\bigskip

\noindent  1. The phases are separated by four coexistence curves. Two coexistence curves coincide with the submanifolds that separate fibers of different types. In particular,  the surface $u_R = u_{max}(\xi_0)$ separates between fibers of type I and fibers of type II, and the surface $u_R = u_f(\xi_0)$ separates between fibers of type II and fibers of type III. {\em Phase transitions across these surfaces are continuous}, because the value of $v_R$ at maximum entropy is continuous. Hence, the temperature at infinity $T_{\infty}$ is also continuous. The former surface describes the F-S phase transition, and the second surface describes the $B_I$-$B_{II}$ phase transition.

The F-$B_I$ and the F-$B_{II}$ transitions occur within the same fiber. As explained earlier, $v_R$ is discontinuous along the phase transition, so the transitions are of first order.

There is one triple point (actually a curve in the thermodynamic state space $Q$) for the F-$B_I$-$B_{II}$ phases. Both $u_R$ and $\xi_0$ on the triple point decrease with $R$.

\bigskip

 \noindent 2. There is no coexistence curve between the S-  and either of the $B_I$ and the $B_{II}$ phases. The S phase has only a coexistence curve with the R phase. This leads to a rather `bizarre' behavior, of a thin strip of F-phase being intermediate between the $B_{II}$ and the S phase. However, this is mathematically necessary, since one cannot go from positive values of $m_0$ to negative values of $m_0$ without crossing the surface $m_0 = 0$. In absence of this strip, the $F$ phase is always at lower energy than the black hole phases, in accordance with the Page-Hawking phase transition \cite{HaPa} or the heuristic discussions of black hole formation in a box \cite{Davies, Page2}.

The strip of F-phase may be removed, if the Bekenstein-Hawking expression for entropy changes at small masses. A small-mass black hole emits more energy in Hawking radiation, and in presence of the box the black hole would have to coexist with its Hawking radiation. If the Hawking radiation contributes negatively to the energy, then it would be possible to have $m_o < 0$ even in presence of the horizon. It would then be possible to pass from the S phase to the B phase without crossing from the R phase. However, such a modification is not only conjectural\footnote{
Note that the usual logarithmic corrections to the Bekenstein-Hawking entropy---see, for example  Ref. \cite{DMB02}---apply in the regime of large masses, and they are not relevant to this problem. A treatment of a black hole in a box with backreaction from its Hawking radiation \cite{AnSav14} leads to a correction linear with respect to mass, but again this works only for black holes of sufficiently large mass. }, but, at the current state of knowledge, it can only be implemented by inserting by hand a phenomenological parameter in the Bekenstein-Hawking formula.

\bigskip

\noindent 3. In linear scale for $u_R$, the F-phase can only be distinguished in the thin strip interpolating between the $S$ and $B_{II}$ phases. We need a logarithmic scale for $u_R$ in order to see the intuitively obvious result that the F-phase dominates at small masses.

  The B-phases dominate at high $R$ and they are suppressed at small $R$. This is obvious since the horizon contribution to the entropy grows faster than any other contribution with the scale of the system. However,  the B-phases disappear as  $\xi_0 \rightarrow -\infty$ for all $R$, reflecting the fact that there are no horizons in a {\em ball} of self-gravitating radiation.

\bigskip

\noindent 4.  $B_I$ solutions have smaller ADM mass than $B_{II}$ solutions for the same $\xi_0$ and $R$. However, the area of the horizon (determined by $m_0$) in $B_I$ is typically larger, because a large part of the mass of $B_{II}$ consists of radiation in the shell.

Two different black hole phases, one with little surrounding radiation and a large horizon , and one with a small horizon also appear in the heuristic analysis of a black hole coexisting with radiation in a box \cite{Davies}. The physical systems under consideration are different, but the main distinction is that in the model of \cite{Davies}, the large black hole phase is unstable, while here the phase $B_I$ is stable.

\subsection{The entropy function in equilibrium}
\begin{figure}[H]
    \centering
   {{\includegraphics[width=15cm]{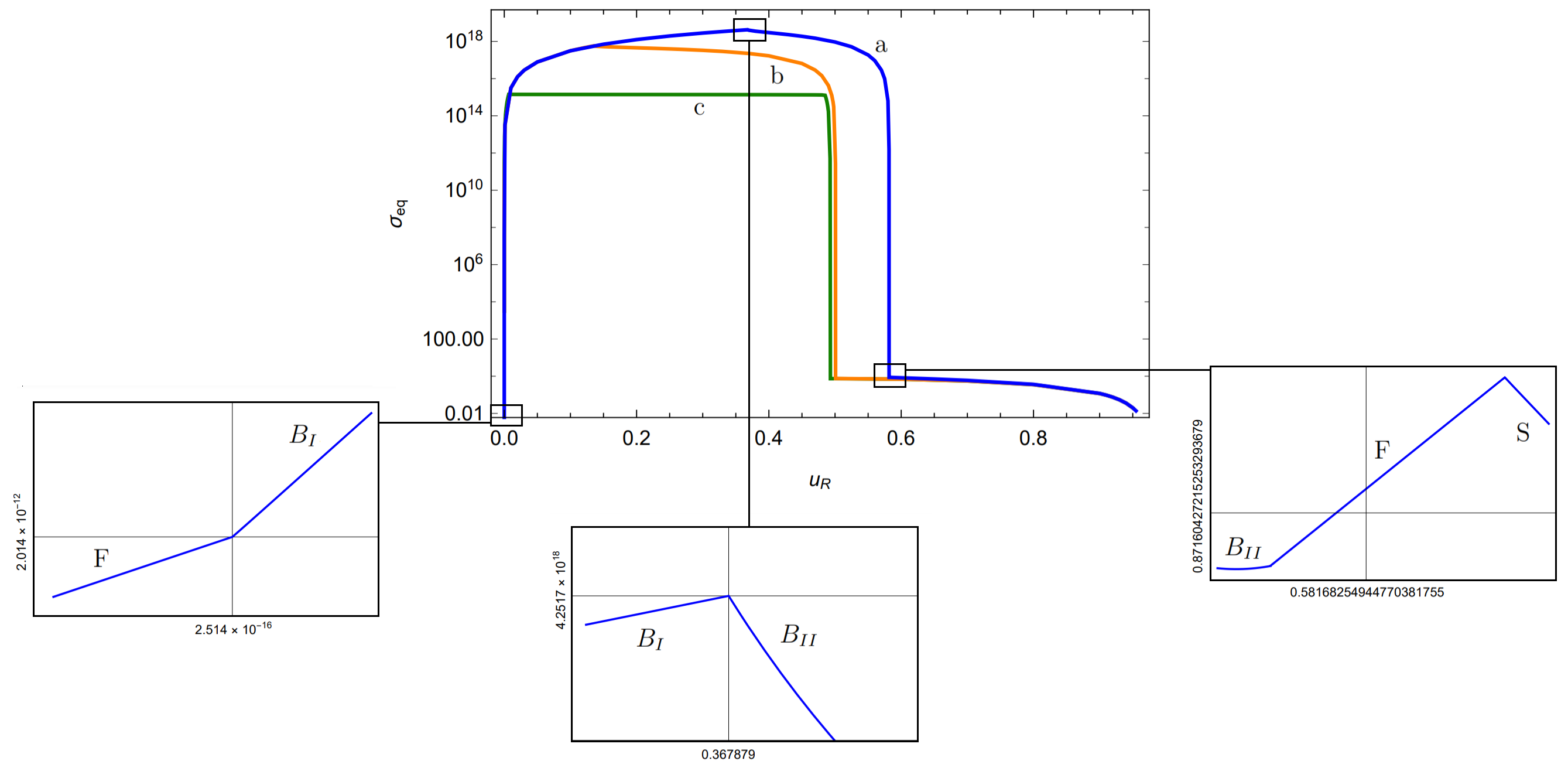} }}
  \caption{Plot of $S_{eq}/R^{3/2}$ as a function of $u_R$ for  $R=10^{38}$  and (a) $ \xi_0=-1$, (b) $\xi_0=-2$, (c) $\xi_0=-5 $. For $\xi_0 = -1$ auxiliary graphs  zoom in specific ranges of $u_R$ where the transitions F- $B_I$, $B_I - B_{II}$ and $B_{II}$-F-S take place, respectively. }
  \label{fig:sequR}
\end{figure}

After the implementation of the MEP, we evaluate the entropy function $S_{eq}(R, u_R, \xi_0)$ on $Q$, using Eq. (\ref{Sz}). Characteristic plots of $S_{eq}$ as a function of $u_R$, for fixed $\xi_0$ and $R$ are given in Fig. \ref{fig:sequR}.

We see that the entropy function is, in general, a non-concave function of $u_R$, and hence, of $M$. By construction, $S$ is continuous across phase transitions.

We note that the entropy is not an increasing function of $u_R$, and that it is bounded from above for fixed $R$ and $\xi_0$. This maximum satisfies Bekenstein's bound \cite{Bek3, Bek4}, $S< 2\pi M R$, or equivalently
\bey
\frac{S}{u_R R^2} < \pi,
\eey
for all values of $R$ that can be reasonably be considered as macroscopic.

We emphasize that it is the consistent implementation of the MEP, through the inclusion of the term $S_{sing}$, that makes this system satisfy Bekenstein's bound.

\subsection{Temperature and heat capacity}
The maximization of the MEP also allows us to identify $v_R$ as a function on $Q$ for the equilibrium solutions. Then, the temperature at infinity is
\begin{eqnarray}
T_{\infty}(R, \xi_0, u_R) = \sqrt{1-u_R} \left(\frac{b v_R(R, \xi_0, u_R)}{4 \pi R^2}\right)^{-1/4}
\end{eqnarray}
 The temperature $T_R$ cannot be identified with a partial derivative of the entropy function. This is only possible for solutions with a simply connected boundary \cite{AnSav14}, which includes regular solutions to the TOV equation \cite{PWZ}.

\begin{figure}[H]
    \centering
   {{\includegraphics[width=16cm]{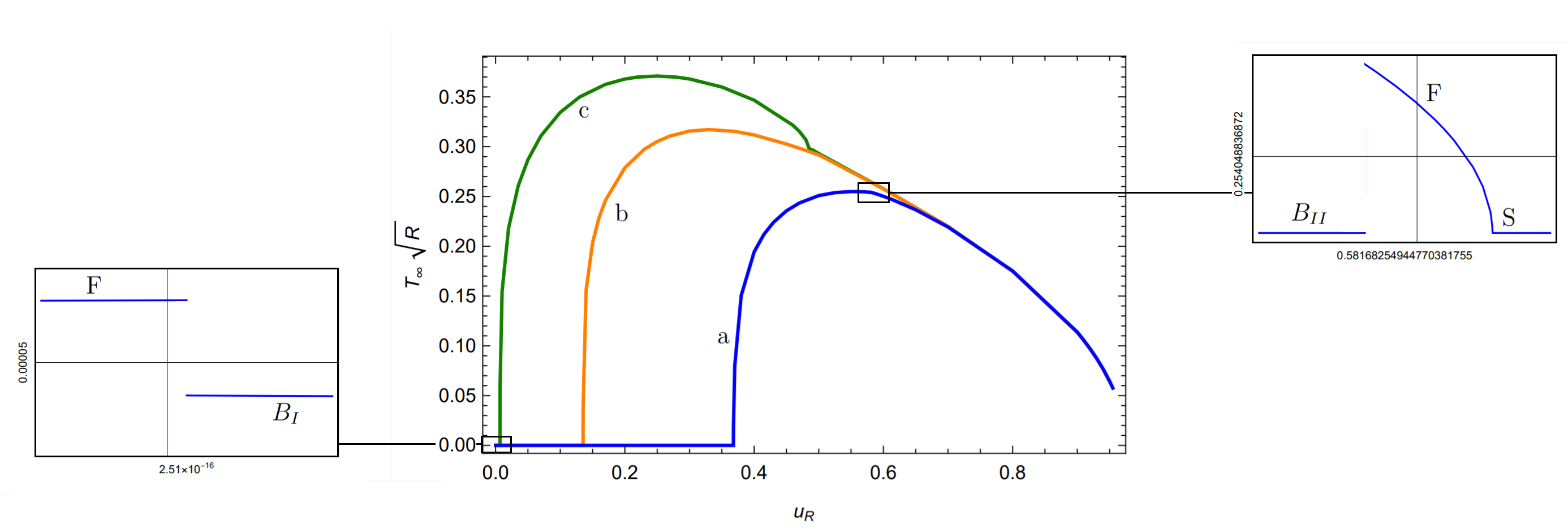} }}
  \caption{Plot of $T_R$ as a function of $u_R$ for fixed $R=10^{38}$ and different values of $\xi_0$:(a) $ \xi_0=-1$, (b) $\xi_0=-2$, (c) $\xi_0=-5 $. The insets zoom near the phase transition points for curve (a) and demonstrate a finite discontinuity for $T_R$.}
  \label{fig:1latcv}
\end{figure}

 Plots of $T_R$ as a function of $u_R$, for fixed $\xi_0$ and $R$ are given in Fig. \ref{fig:1latcv}. The temperature $T_R$ is not an increasing function of the total energy $M$ for fixed $R$ and $r_0$, and it has finite jumps at first-order phase transitions.

The natural definition of heat capacity $C$ when the boundaries of the system are kept fixed (the analogue of heat capacity at constant volume) is \cite{PL}

\begin{equation}
    C:=\left(\frac{\partial M}{\partial T_{\infty}}\right)_{R, r_0} = \frac{R}{2}  \left(\frac{\partial T_{\infty}}{\partial u_R}\right)_{R, \xi_0}^{-1}.
\end{equation}

Clearly, the heat capacity is negative in any region of the thermodynamic state space $Q$ where $T_R$ decreases with $u_R$---see,  Fig. \ref{fig:1latcv}. At the local maxima of $T_R$,  $\left(\frac{\partial T_{\infty}}{\partial u_R}\right)_{R, \xi_0} = 0$, hence, $C$ diverges and changes sign. At the points of the first-order phase transition,  $C$ exhibits discontinuities. Similar to the ball of self-gravitating radiation that was studied in Ref. \cite{PL}, the present system is also characterized    by alternating regions  of positive and negative heat capacities throughout the thermodynamic state space.

\subsection{Latent heat}

Since the transitions F-$B_I$ and $B_{II}$-F are first-order, they involve latent heat. In extensive systems, the latent heat $L$ is identified as the difference $\Delta H$ of the enthalpy between the two phases, while the Gibbs free energy is constant. Hence, $L = \Delta H = \Delta (G +TS) = \Delta(TS)$. In the fundamental representation, $S$ is constant along the transition, hence,    $L = S \Delta T$.

We employ the analogue of this formula in our system, i.e., we consider the quantity
\begin{eqnarray}
L = S \Delta T_{\infty} \label{latent}
\end{eqnarray}
as a candidate for the latent heat in the first order transitions between flat space and the black hole phases. This choice is based primarily on the basis of analogy with extensive thermodynamics. However, it appears plausible that this is the amount of heat (\ref{latent}) we must give the system on the coexistence curve---together with work $\Delta W = - L$, so that $\Delta U = 0$---for the phase change to occur.

\begin{figure}[H]
    \centering
   {{\includegraphics[width=13cm]{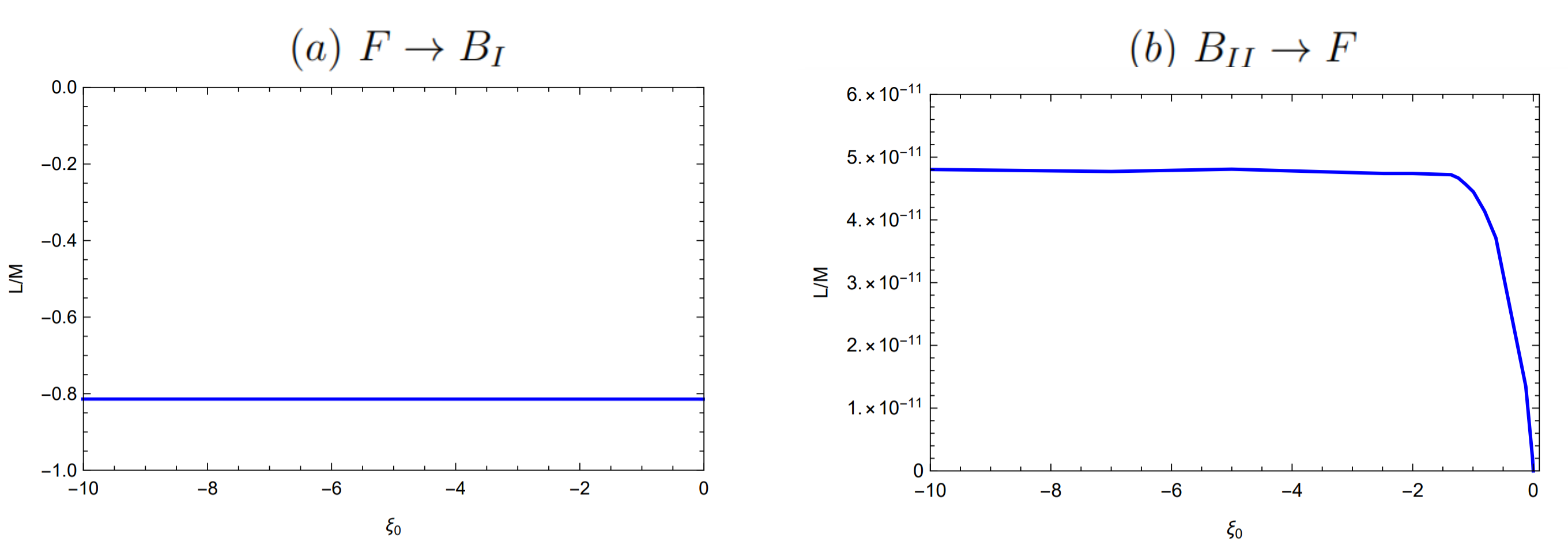} }}
  \caption{The reduced latent heat $L/M$ as a function for $\xi_0$ and for constant $R$: (a)   $F\to B_{I}$ transition, (b)$B_{II} \to F$ transition. }
  \label{fig:latBtoR}
\end{figure}

In Fig. (\ref{fig:latBtoR}), we plot the reduced latent heat $L/M$ as a function of $\xi_0$, for fixed $R$. We note the following.

\medskip

\noindent 1. The latent heat is   negative  for both the $F\to B_{I}$ and the $F\to B_{II}$ transitions.
This means that the black hole phases are always low temperature phases compared to flat space, one needs to `boil' a black hole in order to remove the horizon. This result may appear surprising. Except for the  the narrow strip of F phase before the S phase, the F phase has lower internal energy than the B phases. However, the temperature of the F phase is higher.  This is due to the fact that the temperature $T_R$ is not a monotonous function of the energy.

To the best of our knowledge, the fact that   black holes have lower temperature than self-gravitating systems along the coexistence curve has not been noted. It is easy to see that this must be the case. Consider, for example, a simplified analysis of black hole phase transitions in the vein of \cite{Davies}. A box of radius $R$ contains either a black hole of  mass $M$ or (non-gravitating) radiation with the same mass $M$. In the former case, the entropy is given by the Bekenstein-Hawking expression $S_1 = 4 \pi M^2$, in the latter by $S_2 = \frac{4 \pi}{3} b^{1/4} M^{3/4} R^{3/4}$, where $b$ is the constant in Eq. (\ref{eos}). The associated temperatures are $T_1 = (8\pi M)^{-1}$ and $T_2 = \frac{1}{\pi} b^{-1/4} M^{1/4} R^{-3/4}$.

A black hole is entropically favored for $M > M_c$, where the critical mass  $M_c = \left(\frac{b}{81}\right)^{1/5} R^{3/5}$. The temperature of the black hole phase at $M_c$ is $T_1(M_c) = \frac{1}{8 \pi} \left(\frac{81}{b}\right)^{1/5} R^{-3/5}$/ The temperature of the radiation phase is $T_2(M_c) = \frac{1}{\pi(81)^{\frac{1}{20}}} b^{-1/5} R^{-3/5}$. We compute $\Delta T = T_1(M_c) - T_2(M_c)$,
\bey
\Delta T = -\frac{5}{24 \pi R^{3/5}} \left(\frac{81}{b}\right)^{1/5} < 0,
\eey
i.e., we verify that the black hole has lower temperature than radiation along the coexistence curve.

It is important to understand the role  of the negative latent heat in a non-equilibrium setting of gravitational collapse. A plausible conjecture is that it refers to the amount of radiation (electromagnetic or gravitational) that must be emitted before the  system settles as a black hole. To test this conjecture, we must undertake an analysis of gravitational collapse in the context of non-equilibrium thermodynamics, while keeping track of all heat currents throughout the collapse.

\medskip
\noindent 2. The reduced latent heat is almost constant for a large range of values of $\xi_0$. It vanishes as $\xi_0 \rightarrow - \infty$, well outside the range of the plots in Fig. (\ref{fig:latBtoR}), because the black hole phases disappear at this limit.

\medskip

\noindent 3. The latent heat is much smaller in transitions involving the $B_{II}$ phase than in transitions with the $B_I$ phase. This is because the horizon in the $B_{II}$ phase is much smaller. In contrast, the latent heat with respect to the $B_I$ phase is a substantial fraction of the total mass.

\section{Conclusions}

We analysed the thermodynamics of a shell of gravitating radiation surrounding a solution to vacuum Einstein's equation, which may either correspond to flat space, a black hole or a repulsive singularity. The shell can be interpreted as a   gravitating heat reservoir. However, the presence of long range forces necessitates an analysis of the total system that consists of the shell and its interior.

We showed that the only way to obtain a consistent thermodynamic description of the system is by assigning a specific expression for entropy to the naked singularity, thus, confirming the proposal of Ref. \cite{AnSav12} in a more complex set-up. The result is a concrete model for describing phase transitions between black holes and self-gravitating systems that is fully compatible with the rules of thermodynamics. Such models are important for black hole thermodynamics and quantum gravity, but also for expounding the mathematical and physical structure of non-extensive gravitating systems.

The methods of this paper can be straightforwardly generalized, for example, to different equations of state, rotating systems and other shell geometries. The self-gravitating shell of radiation  can also be used as a generic thermal reservoir in studies of system-reservoir thermodynamics in self-gravitating systems.

Furthermore, our results strongly suggest the importance of a thermodynamic analysis to gravitational collapse. The non-equilibrium evolution of a self-gravitating shell is perhaps the simplest model for analyzing the interplay between horizon formation and thermodynamics in a gravitating system. We expect that the solutions studied in this paper  will correspond to asymptotic states of a non-equilibrium analysis.

\section*{Acknowledgements}

D.K. acknowledges financial support from the  “Andreas  Mentzelopoulos Foundation”.

\appendix

\section{The asymptotic behavior of radiation entropy}
In this section, we prove the asymptotic behavior of the radiation entropy $S_{rad}(u_R, v_R, R, \xi_0)$ of (\ref{srad0}), as $v_R\rightarrow \infty$ and as $v_R \rightarrow 0$, the other parameters being fixed. In particular, we show that (i) $S_{rad}\rightarrow \infty$ as $v_R \rightarrow \infty$, for all $u_R, R, \xi_0$, (ii)  $S_{rad}\rightarrow \infty$ as $v_R \rightarrow 0$, for $\xi_0 < \log u_R$, and (iii)$S_{rad}\rightarrow 0$ as $v_R \rightarrow 0$, for $\xi_0 >\log u_R$.

\subsection{Main formulas}
The function $S_{rad}$ can be expressed as
\bey
S_{rad} = \frac{2}{9}(4\pi b )^{1/4} [\sigma(0) - \sigma(\xi_0)]R^{3/2},
\eey
where
\bey
\sigma(\xi) := \frac{u(\xi)+\frac{3}{2}v(\xi)}{v(\xi)^{1/4}\sqrt{1-u(\xi)}}e^{3\xi/2}.
\eey
The functions $u(\xi), v(\xi)$ are solutions to the differential equations
\begin{equation}
    u' = 2v-u, \qquad\qquad v' = \frac{2v(1-2u-\frac{2}{3}v)}{1-u}, \label{evoleq}
\end{equation}
 for $\xi < 0$ and with boundary conditions $u(0) = u_R$ and $v(0) = v_R$.

Using Eqs. (\ref{evoleq}), we find
\begin{equation}
    \frac{d\sigma}{d\xi} =  \frac{6 v^{3/4}e^{3\xi/2}}{\sqrt{1-u}} \geq 0. \label{dedkx}
\end{equation}
Hence, $\sigma(0) - \sigma(\xi) \geq 0 $, for all $\xi \leq 0$.

It is convenient to introduce the variable $\epsilon:= 1 - u$, bringing Eqs. (\ref{evoleq}) into the form
\begin{equation}
    \epsilon' = 1-\epsilon - 2 v,  \qquad \qquad v' =\frac{2v(2\epsilon-1-\frac{2}{3}v)}{\epsilon}. \label{evoleq2}
\end{equation}

An important special case is the regime where either $v >> 1$ or $\epsilon >> 1$. In this regime, Eqs. (\ref{evoleq2}) become
\bey
\epsilon' = -\epsilon - 2 v,  \qquad \qquad v' =\frac{2v(2\epsilon-\frac{2}{3}v)}{\epsilon}, \label{evoleq3}
\eey
and they admit exact solution
\bey
\frac{\epsilon}{v} = \alpha_1 e^{-5\xi} - \frac{2}{15}, \qquad
\epsilon (\xi) = \alpha_2 e^{-\xi}\left(\frac{15}{2} \alpha_1 - e^{5\xi}\right)^3, \label{solapp1}
\eey
where $\alpha_1$ and $\alpha_2$ are integration constants.
For these solutions, Eq. (\ref{dedkx}) becomes
\bey
   \frac{d\sigma}{d\xi} = 6\left(\frac{15}{2}\right)^{3/4} \alpha_2^{1/4} e^{5 \xi}. \label{dsdks}
\eey

\subsection{Case $v_R \rightarrow \infty$}
Consider the case of $v_R >> 1$.   Eq. (\ref{evoleq3}) applies in the vicinity of $\xi = 0$. As shown in \cite{AnSav21}, in all singular solutions, $v$ increases with decreasing $\xi$ up to a point $\xi_2$, where $v'(\xi_2) = 0$, and then it decreases to zero as $\xi \rightarrow -\infty$. However, at $\xi = \xi_2$, $v \simeq 3 \epsilon$, hence, $\epsilon(\xi_2) >> 1$. For $\xi < \xi_2$, $\epsilon$ keeps increasing with decreasing $\xi$.

Hence, Eq.  (\ref{evoleq3}) applies to all $\xi \leq 0$, and we can evaluate the integration constants in Eq. (\ref{solapp1}) by the values of $\epsilon$ and $v$ at $\xi = 0$. Then, we obtain
\bey
\alpha_1 = \frac{2}{15} + \frac{\epsilon_R}{v_R}, \qquad     \alpha_2 = \bigg(\frac{2}{15}\bigg)^3\frac{v_R^3}{\epsilon_R^2}.
\eey
Hence, Eq. (\ref{dsdks}) gives
\bey
  \frac{d\sigma}{d\xi} = \frac{v_R^{3/4}}{\epsilon_R^{1/2}} e^{5\xi}.
\eey
Thus, we obtain
\bey
\sigma(0) - \sigma(\xi) =  \frac{6}{5}\frac{v_R^{3/4}}{\sqrt{1-u_R}}(1-e^{5\xi_0}),
\eey
to conclude that $\lim_{v_R \rightarrow \infty}[\sigma(0) - \sigma(\xi)] = \infty$.

\subsection{Case $v_R \rightarrow 0$ and $\xi_0 > \log u_R$}

Consider a solution with  $v_R << u_R < 1$. Integrating from $\xi = 0$,  $u$ increases with decreasing $\xi$; $v$ initially decreases with decreasing $\xi$ and then increases again. The condition $v<< u$ remains valid for an interval $(\xi_r, 0)$, in which
\begin{eqnarray}
u' = - u, \qquad
v' = \frac{2v (1-2u)}{1-u}. \label{du12}
\end{eqnarray}
These equations admit solutions
\begin{eqnarray}
 u(\xi) = u_R e^{-\xi}, \qquad   v = \frac{v_R u_R^2 (1 - u_R)^2}{u^2 (1-u)^2} \label{vx1}
\end{eqnarray}
For sufficiently small $v_R$, Eq. (\ref{vx1}) applies up to a point where  $\epsilon = 1 - u <<1$.

 For $\epsilon << 1$ (but not necessarily $v<< 1$), Eqs. (\ref{evoleq2}) become
\begin{eqnarray}
\epsilon' = 1 - 2v, \qquad
v' = -\frac{2v (1 +\frac{2}{3}v)}{\epsilon}, \label{de2}
\end{eqnarray}
Hence, we obtain
\begin{eqnarray}
\frac{d \epsilon}{dv} = -\frac{\epsilon (1-2v)}{2v (1 +\frac{2}{3}v)} \label{veb}
\end{eqnarray}
Eq. (\ref{veb}) has solutions of the form
\begin{eqnarray}
\frac{v}{(v+\frac{3}{2})^4} = \frac{a}{\epsilon^2}, \label{vu1a}
\end{eqnarray}
for some constant $a$.

The minimum value $\epsilon_*$ of $\epsilon$ occurs at $\xi = \xi_*$, such that $\epsilon'(\xi_*) = 0 $, or equivalently $v(\xi_*) = \frac{1}{2}$.
 By   Eq. (\ref{vu1a}),  $a = \epsilon_*^2/32$. Then, Eq. (\ref{vu1a}) becomes
\begin{eqnarray}
\frac{32v}{(v+\frac{3}{2})^4} = \left(\frac{\epsilon_*}{\epsilon}\right)^2. \label{vu2}
\end{eqnarray}

Eqs. (\ref{vx1}) and (\ref{vu2}) must approximately coincide in some open set of $\xi$ where  $v<< 1$. This is only possible if
\begin{eqnarray}
\epsilon_* = \frac{16}{9} u_R (1 - u_R) \sqrt{2 v_R}. \label{e0b}
\end{eqnarray}

  Eqs. (\ref{vu2}) and (\ref{de2}) imply that
\begin{eqnarray}
(v^{-1/2} + \frac{3}{2} v^{-3/2}) v' = - \frac{16 \sqrt{2}}{3 \epsilon_*}.
\end{eqnarray}
Integrating from some reference point $\xi = \xi_r$ with $v(\xi_r) = v_r$, we find
\begin{eqnarray}
2(\sqrt{v(\xi)} - \sqrt{v_r}) - 3 \left(\frac{1}{\sqrt{v(\xi)}} - \frac{1}{\sqrt{v_r}}\right) =  - \frac{16 \sqrt{2}}{3 \epsilon_*} (\xi - \xi_r) \label{vx2}
\end{eqnarray}

Using Eq. (\ref{vx2}) for a choice of the reference point $\xi = \xi_r$ lying in the domain of validity of Eq. (\ref{vx1}),
\begin{eqnarray}
\xi = \log u_R + \frac{3 \epsilon_*}{16 \sqrt{2}} \left( \frac{3}{\sqrt{v(\xi)}} - 2 \sqrt{v(\xi)} \right). \label{xv}
\end{eqnarray}
Setting $\xi = \xi_*$ in Eq. (\ref{xv}), we obtain
\begin{eqnarray}
\xi_* = \log u_R + \frac{3 \epsilon_*}{8}. \label{xv1}
\end{eqnarray}

The key point in this analysis is that $\epsilon_*\rightarrow 0$ for $v_R \rightarrow 0$. By Eq. (\ref{xv1}), the smallest value for $\xi_0 < \xi_*$ is $\log u_R$. For any $\xi_0 < \log u_R$, we can choose sufficiently small $v_R$, so that $\frac{\xi_0 - \log u_R}{\epsilon_*} >> 1$, which by Eq. (\ref{xv}) implies that $v(\xi_0) << 1$. Hence, for sufficiently small $v_R$, $\xi_0$ lies always in the domain of validity of Eq. (\ref{vx1}). Then, Eq.
(\ref{dedkx}) becomes
\bey
    \frac{d\sigma}{d\xi} = \frac{6 v_R^{3/4}u_R^{3/2}(1-u_R)^{3/2}}{(1-u_Re^{-\xi})^2}e^{3\xi}.
\eey
We integrate to obtain
\bey
\sigma(0) - \sigma(\xi_0) = 6 v_R^{3/4}u_R^{3/2}(1-u_R)^{3/2} F(\xi_0),
\eey
where $F(\xi_0) = \int_0^{\xi_0} \frac{d\xi e^{3\xi}}{(1 - u_Re^{-\xi})^2}$ is a smooth function of $\xi_0  \in (\log u_R, 0)$.
We conclude that $\lim_{v_R \rightarrow 0}[\sigma(0) - \sigma(\xi)] = 0$.


\subsection{Case $v_R \rightarrow 0$ and $\xi_0 < \log u_R$}
As shown in Ref. \cite{AnSav21}, the solution  $\xi < \xi_*$ is characterized by a point $\xi_1$, such that $u(\xi_1) = 0$, or $\epsilon (\xi_1) = 1$. By continuity, there exists an interval  $(\bar{\xi}, \xi_*)$ where $ \xi_* >\bar{\xi} > \xi_1  \epsilon(\xi) $ remains smaller than any arbitrary value $1 > \bar{\epsilon} > \epsilon_*$. Since $\epsilon_* \sim \sqrt{v_R}$, we can choose $\bar{\epsilon}$ to be proportional to $v_R^a$, for $a < \frac{1}{2}$. It is convenient to choose
\bey
\bar{\epsilon} = \frac{u_R(1-u_R)}{9\sqrt{2}} v_R^{1/4},
\eey
 so that $\bar{v} = v(\bar{\xi}) = v_R^{-1/6}$.

Eq. (\ref{vx2}) for   $\xi_r = \xi_*$ becomes
\bey
( 2 \sqrt{v(\xi)} - \frac{3}{\sqrt{v(\xi)}} - 2 \sqrt{2} ) = \frac{16}{\sqrt{2}\epsilon_*}(\log u_R - \xi). \label{vx4}
\eey
For fixed $\xi < \log u_R$, we can choose $v_R$ so that the right hand side of Eq. (\ref{vx4}) becomes very large. Since $v > \frac{1}{2}$ for $\xi > \xi_*$, in this limit $v>> 1$, hence,
\bey
v(\xi) = \frac{32}{\epsilon_*^2} (\log u_R - \xi)^2.
\eey
In this regime, Eq. (\ref{vu2}) implies that $\epsilon = \epsilon_* \frac{v^{3/2}}{4\sqrt{2}}$. Then, Eq.
(\ref{dedkx}) becomes
\bey
\frac{d\sigma}{d\xi} = 2^{1/4} \frac{12 }{\sqrt{\epsilon_*}}.
\eey
Hence, integrating from $\log u_R$ to $\xi_0 \in (\log u_R, \bar{\xi})$, we find
\bey
\sigma(\log u_R) - \sigma(\xi_0)  = 2^{1/4} \frac{12 }{\sqrt{\epsilon_*}} (\log u_R - \xi_0).
\eey
We see that $\lim_{v_R \rightarrow 0}
\sigma(\log u_R) - \sigma(\xi_0) = \infty$. Since $\sigma(0)  - \sigma(\xi_0) > \sigma(\log u_R) - \sigma(\xi_0) $, we conclude that for $\xi_0 \in (\bar{\xi}, \log u_R)$, $\lim_{v_R \rightarrow 0}
\sigma(0) - \sigma(\xi_0) = \infty$.

Finally, we examine the case of $\xi < \bar{\xi}$. Since $\bar{v}\rightarrow \infty$ for $v_R \rightarrow 0$, the solution  from $\bar{\xi}$ to the center is of the type that has been studied in Sec. A.2. Hence, for any $\xi_0 < \bar{\xi}$,
$\lim_{v_R \rightarrow 0}
[\sigma(\bar{\xi}) - \sigma(\xi_0)] = \infty$. Since $\sigma(0)  - \sigma(\xi_0) > \sigma(\log u_R) - \sigma(\xi_0) $, we conclude that for $\xi_0 \in (-\infty, \bar{\xi}]$, $\lim_{v_R \rightarrow 0}
[\sigma(0) - \sigma(\xi_0)] = \infty$.

\end{document}